\documentclass[%
 reprint,
preprintnumbers,
 amsmath,amssymb,
 aps,
prd,
]{revtex4-2}

\usepackage[utf8]{inputenc}
\usepackage[T1]{fontenc}
\usepackage[bookmarks=true, bookmarksnumbered=true, colorlinks=true, citecolor=blue, linkcolor=blue, breaklinks=true]{hyperref}

\usepackage{color,graphicx}
\usepackage{siunitx}
\usepackage{bm}
\usepackage{longtable, multirow}

\usepackage{comment}

\usepackage{orcidlink}

\definecolor{nicegreen}{rgb}{0.1,0.5,0.1}
\definecolor{darkblue}{rgb}{0.15, 0.2, .85}
\definecolor{darkgreen}{rgb}{0.1,0,0.3}
\definecolor{darkred}{rgb}{0.6,0,0}
\newcommand*{\B}[1]{\ifmmode\bm{#1}\else\textbf{#1}\fi}

\newcommand{\dyne}{\, \rm dyne}

\newcommand{\km}{\, \rm km}
\newcommand{\cm}{\, \rm cm}
\newcommand{\g}{\, \rm g}
\newcommand{\s}{\, \rm s}
\newcommand{\Hz}{\, \rm Hz}

\newcommand{\nsat}{n_\text{sat}}

\begin{document}

\hfill \preprint{KCL-PH-TH/2024-58}

\hfill \preprint{FERMILAB-PUB-24-0778-T}

\title{Direct nonparametric multimessenger constraints on the equation
of state of cold dense nuclear matter}

\author{Iuliu Cuceu~\orcidlink{0009-0005-5243-3347}}
\email{icuceu@oca.eu}
\affiliation{Department of Physics, King’s College London, Strand, London, WC2R 2LS, UK}
\affiliation{Physics Department, Universit\`{a} di Roma ``Tor Vergata'', I-00133 Roma, Italy}
\affiliation{Department of Astronomy, Faculty of Mathematics, University of Belgrade, Studentski trg 16, 11000 Belgrade,
Serbia}

\author{Sandra Robles~\orcidlink{0000-0002-6046-8217}}
\email{srobles@fnal.gov}
\affiliation{Theoretical Particle Physics and Cosmology Group, Department of Physics, King’s College London, Strand, London, WC2R 2LS, UK}
\affiliation{Astrophysics Theory Department, Theory Division, Fermi National Accelerator Laboratory, Batavia, Illinois 60510, USA}

\date{\today}

\begin{abstract}

We utilize the now substantial amount of astrophysical observations of neutron stars (NSs), along with
perturbative quantum chromodynamics (pQCD) calculations at high density, to directly constrain the NS
equation of state (EOS). To this end, we construct nonparametric EOS priors by using Gaussian processes
trained on 75 EOSs, which include models with either hadrons, hyperons, or quarks at high densities. We
create a prior using the full EOS sample (model agnostic), and one prior for each EOS family to test model
discrimination. We introduce a novel inference approach, which allows the simultaneous sampling of
intrinsic and extrinsic parameters of binary NS mergers, as well as a nonparametric equation of state. We
showcase this method in a Bayesian updating scheme by first performing a complete analysis of the binary
NS merger event GW170817 with minimal assumptions, and sequentially adding information from x-ray
and radio NS observations, along with pQCD calculations. Besides providing standard constraints, such as
the pressure at twice nuclear saturation density
$p(2\rho_\text{sat})=4.3^{+0.6}_{-0.6}\,\times 10^{34}\text{dyne/cm}^{2}$, at $95\%$ confidence
level, for the model agnostic prior, our methodology shows how the choice of EOS families used in
conditioning changes the inferred astrophysical properties of the EOS, namely tidal deformability and
maximum supported NS mass. We find hyperonic priors predicting higher tidal deformabilities for a $1.4M_\odot$ NS, and hadronic priors being preferred by the considered astrophysical data. 

\end{abstract}

\maketitle

\section{Introduction}

The study of neutron stars (NSs) offers novel and unique insights into the behavior of extremely dense cold nuclear matter, which cannot be replicated under terrestrial laboratory conditions. Neutron stars, the remnants of massive stellar progenitors, compress $\sim1-2$ solar masses into a radius of about $10-12\,\km$~\cite{Ozel:2016oaf}, reaching densities that far exceed nuclear saturation density. Understanding their internal structure and the equation of state (EoS) governing such extreme environments has profound implications for both nuclear physics and astrophysics~\cite{Schaffner-Bielich_2020}. Besides the lack of direct experimental data of matter in the NS density regime, another challenge is the intrinsic complexity of modeling interactions among nucleons and potentially more exotic particles like hyperons and quarks at such high densities.

The core of a NS is an extreme environment, with average densities above the nuclear saturation mass density $\rho_\text{sat}\simeq 2.7\times10^{14}\g\cm^{-3}$~\cite{Ozel:2015fia}. 
This high density gives rise to unique macroscopic effects, such as rapid rotation and intense magnetic fields that can be observed, thereby providing insights into the microphysics of the NS interior. In order to determine the macroscopic properties of NSs by solving the Tolman-Oppenheimer-Volkoff (TOV) equations~\cite{Tolman:1939jz, Oppenheimer:1939ne}, such as masses and radii, an EoS is needed to model the various layers of the star and the transitions between them.

Traditional approaches to modeling the EoS have relied on parametric models such as the piecewise polytrope~\cite{Lindblom:2018rfr} or spectral decomposition methods~\cite{Lindblom:2010bb}. These techniques allow for computational efficiency and have been widely used to infer NS properties from observational data~\cite{Read:2008iy,Lindblom:2012zi,Lindblom:2013kra,Raaijmakers:2019dks,Jiang:2022tps}. However, they often suffer from significant limitations, particularly when modeling phase transitions or the emergence of new particle species at high densities~\cite{Carney:2018sdv}. Parametric models inherently introduce biases due to their fixed functional forms, making it difficult to capture the full range of possible EoSs~\cite{Legred:2021hdx,Legred:2022pyp}. Moreover, from a methodological standpoint, current observations of NS macroscopic observables have to be analyzed in a self-consistent and model-independent manner. 

In this paper, we aim to overcome these limitations by employing a non-parametric approach using Gaussian processes (GPs)~\cite{Landry:2019} to model the EoS. By training a GP on a wide variety of equations of state, we can construct a prior that is flexible and agnostic to specific parametrizations, thereby allowing the data to drive the inference of the EoS without strong model assumptions. The GP-based method provides a robust and transparent way to capture the wide range of physical behaviors expected from different nuclear matter compositions. Furthermore, we introduce a novel analysis methodology by constructing a categorical distribution over the Gaussian process prior draws, and utilizing Bayesian updating on three different datasets. This
not only allows the non-parametric prior construction direct access to observational data,
but also provides a scalable treatment of an otherwise high-complex prior construction.  To showcase the flexibility of our method, we first run a full parameter inference scheme 
utilizing our prior on the gravitational waveform of the binary neutron star (BNS) merger  event GW170817~\cite{LIGOScientific:2017vwq}. The next Bayesian updating step utilizes the posteriors of 105 NS mass measurements, as well as two  mass-radius posteriors~\cite{Miller:2019cac,Miller:2021qha} from the Neutron Star Interior Composition Explorer (NICER). Lastly, we also showcase the effect of including  recent developments in perturbative Quantum Chromodynamics (pQCD) calculations~\cite{Komoltsev:2021jzg}. 

The combination of non-parametric modeling with Bayesian updating and our method, allows us to perform a fully consistent analysis across multiple datasets in a prior transparent and model independent way. We also aim to showcase the effect of each dataset on the direct NS EoS constraints, as well as discuss likelihood construction and prior selection and their hidden assumptions. By integrating information from different datasets, including gravitational wave observations of NS mergers, NICER measurements of NS mass-radius relations and pQCD calculations at high densities, we present a comprehensive analysis that explores the current boundaries of the nuclear EoS and its potential extensions into new physical regimes. 

The paper is organized as follows: Section~\ref{Sec:eos} outlines the construction of a GP-based prior used for inference. Section~\ref{Sec:methods} introduces our novel treatment of this prior in a parametrized Bayesian updating scheme. Section~\ref{Sec:data} describes the details of our full analysis of the GW170817 waveform, as well as the other astrophysical and theoretical data utilized. Our results are presented in Section~\ref{Sec:results}, and a discussion about the impact of the available data on the prior choice and likelihood construction is given in Section~\ref{sec:conclu}.

\section{Modeling of the Equation of State} 
\label{Sec:eos}

There are a number of EoS parametrizations in the literature, the most prevalent of which being  piecewise polytropes, see e.g. ref.~\cite{Lindblom:2018rfr}, and the spectral decomposition parametrization  \cite{Lindblom:2010bb}. Within a standard Bayesian parameter inference scheme, these techniques have been used to directly constrain the NS EoS (see ref.~\cite{Ozel:2016oaf} for a review). However, parametric models, although providing a certain ease of use, have proven inadequate in other regards. Difficulties with phase transitions have been identified multiple times in both the aforementioned parametrizations \cite{Carney:2018sdv,Lindblom:2010bb,Lindblom:2012zi,Lindblom:2013kra}, and although increasing the number of parameters can somewhat alleviate the issue, it quickly becomes computationally unfeasible. 

The state-of-the-art approach to these issues, within a Bayesian inference context, is the use of non-parametric models, namely Gaussian processes, introduced in ref.~\cite{Landry:2019}. We shall briefly summarize how a GP prior is constructed by training on tabulated equations of state. 

\begin{figure}
    \centering
    \includegraphics[width=\columnwidth]{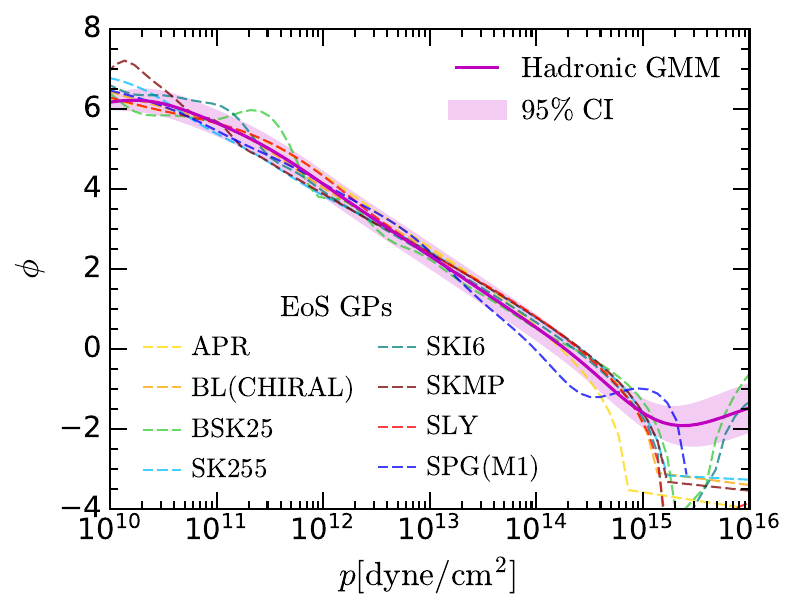}
    \caption{Construction of a Gaussian mixture model (magenta) based on 32 different hadronic unified equations of state. 
    For reference, some  EoSs GP means are overlaid, without the covariances, for clarity. Since a GMM is a collection of multivariate normal distributions, we plot the highest likelihood mean and all the covariances within $95\%$ confidence.}
    \label{fig:GMM}
\end{figure}

A Gaussian process is a non-parametric, Bayesian approach to regression that defines a prior distribution over functions, where any finite collection of function values is jointly Gaussian. By conditioning on observed data, GP regression (GPR) computes a posterior distribution over functions, which allows for smooth interpolation and principled uncertainty quantification. GPR is employed twice, first to resample each tabulated EoS and then to construct an over-arching GP trained on multiple EoSs. The resampling step ensures that multiple sets with varying sizes of tabulated EoSs (energy density $\epsilon$, versus pressure $p$) are treated equally, and effectively provides a smooth interpolation of a set $\epsilon_i(p_i)$, equipped with error estimates. GP regression is performed on the auxiliary variable $\phi=\log(c^2\, d\epsilon/dp-1)$ obtained from numeric differentiation of a tabulated EoS. Thus, for an EoS, $\alpha\in\{\alpha\}_A$, out of the training set, we have the Multivariate Gaussian Distribution model
\begin{widetext}
\begin{equation}
    \phi_i^{(\alpha)}\mid \log p_i,\big\{ \log \epsilon^{(\alpha)}_{j^\ast}, \log p^{(\alpha)}_{j^\ast} \big\} \sim \mathcal{N}\bigg(E^{(\alpha)} (\phi_i), \text{Cov}^{(\alpha)} (\phi_i,\phi_j)\bigg),
\end{equation}
\end{widetext}
 where the starred indices (e.g. $\epsilon_{i^\ast}$) indicate data the GP is conditioned upon, as opposed to generic points (e.g. $\epsilon_{i}$) where GPR is used as an interpolator. The covariance kernel in the normal distribution, $\mathcal{N}$, is analytically computed from a radial basis function (RBF) kernel of a GP over $\epsilon_i(p_i)$, with hyper-parameters obtained from GP likelihood optimization. Similarly, the mean $E$ is constructed from the ordinates $\log \epsilon_{j^\ast}$ and a low order polynomial fit. For a step-by-step guide on how a GP is built from tabulated pressure, energy density pairs, see ref.~\cite{Landry:2019}. From the set of $n$ resampled EoSs modeled by GPs, $\phi^{(\alpha)}(\log p)$, the overarching process, $\phi(\log p)$, is constructed:
 \begin{widetext}
 \begin{equation}
     \phi_i \mid \log p_i , \big\{E^{(\alpha=1)}(\phi_{j^\ast}), E^{(\alpha=2)}(\phi_{j^\ast}), ..., E^{(\alpha=n)}(\phi_{j^\ast}), \log p_{j^\ast}\big\} \sim \mathcal{N} \big(\widetilde{E}, \widetilde{\text{Cov}}\big),
 \end{equation}
 \end{widetext}
with the kernel here being a combination of the RBF and white noise kernels, while the mean is constructed from the means of the individual GPs. Instead of normal hyper-parameter optimization, a Gaussian Mixture Model (GMM) of many GPs with different hyper-parameters is then constructed, as described in ref.~\cite{Essick:2019ldf}, effectively allowing for fine-tuned, and intuitive control over how closely the resulting EoSs are informed by the theoretical ones they were trained upon. This tuning is done through the hyper-prior selection of the cross-validation likelihood:
\begin{equation}\label{eq:crossvalid}
    P_{CV} (\{\alpha\}_A\mid \bm{\sigma})=\prod_{a\in A} P(\alpha^{(a)}\mid \{\alpha\}_{A\diagdown a},p,\bm{\sigma}),
\end{equation}
where $\bm{\sigma}$ is the set of hyper-parameters. An illustration of this construction scheme can be found in Figure~\ref{fig:GMM} in the $\phi(p)$ plane, where we take multiple theoretical equations of state and construct a GMM (magenta) from which we can draw to obtain realistic and model informed EoSs to use as a prior. Note that EoS GP means are overlaid, for clarity we do not show the covariances. Above $\sim 10^{15}\dyne/\cm^2$, the GP covariances are very high, which explains the apparent discrepancy between the GMM and GPs. 

\begin{figure*}
    \centering
    \includegraphics[width=0.497\textwidth]{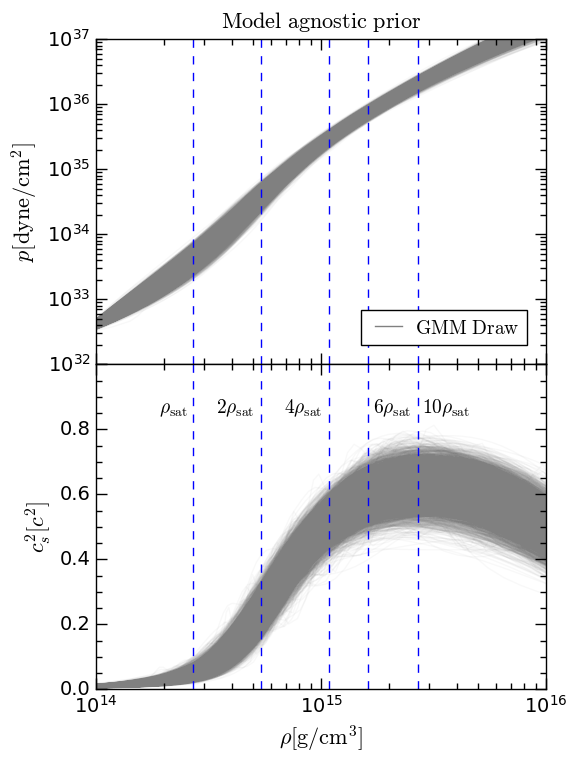}
    \includegraphics[width=0.497\textwidth]{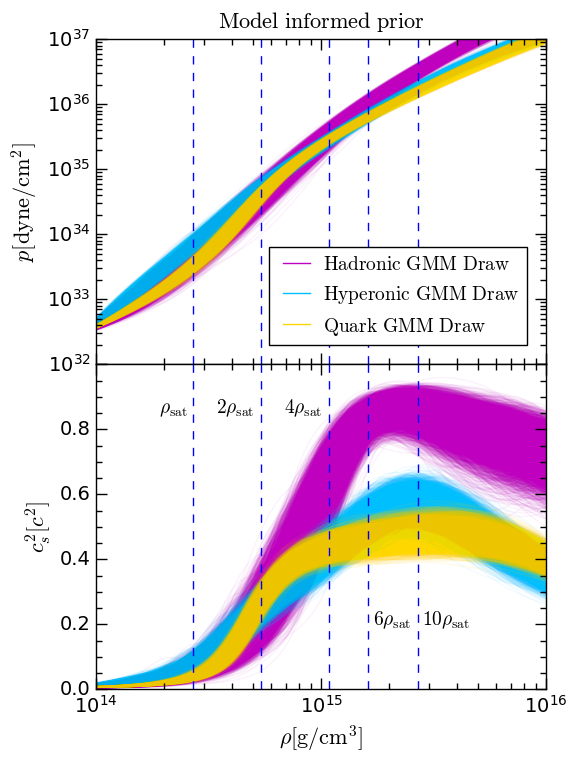}    
    \caption{Construction of the model agnostic and model informed priors. The model agnostic GMM (left) is constructed directly from GPs trained on 75 different EoSs, while the model informed GMMs (right) separates the training set by particle content, with hadronic-trained GMM (magenta), hyperonic-trained GMM (light blue) and quark-trained GMM (yellow). For reference, the vertical dashed blue lines represent the once, twice, four, six and ten times the nuclear saturation density.}
    \label{fig:draws}
\end{figure*}

To showcase both the flexibility of our method and how prior construction more broadly affects the resulting posteriors, we construct two separate priors. The first, called ``model agnostic'', is trained on the 75 EoSs of cold NSs  available in the CompOSE database~\footnote{ \url{https://compose.obspm.fr}} \cite{Typel:2013rza,Oertel:2016bki,CompOSECoreTeam:2022ddl}. Our training  dataset is composed of 
32 hadronic equations of state (baryonic matter composed only of protons and neutrons), 17 hyperonic models (includes baryons with up, down and strange quark content) and 26 quark EoSs that explicitly model a phase transition to deconfined quark matter with varying densities and strengths. 
For more information on the EoSs used and the hyper-priors utilized, see Appendix~\ref{sec:EOSsample}. 
This model agnostic prior is a direct construction of a GMM from the individual GPs trained on each EoS, with wide hyper-priors. The model utilized in Bayesian inference consists of 10000 draws from this GMM construction. In contrast, the second, model-informed EoS prior, has narrower hyper-priors and uses 3 separate GMMs, one for each family of EoSs (hadronic, hyperonic or quark). Each GMM then provides 10000 draws from which we can compute Bayes factors to constrain the likelihood of each family after the Bayesian updating steps. The purpose of this second construction is to showcase how each new data stream and likelihood affects the prior. Unlike the first, it does not aim to provide direct conclusive EoS constraints (since the prior is heavily biased towards a given particle composition of the NS inner core). Both priors can be found in Figure \ref{fig:draws}, in the density-pressure/speed of sound plane, along with reference saturation densities (vertical dashed lines). The GMMs are stitched at very low density crust, $\rho_{\text{stitch}}=10^{10}\g/\cm^3,$ to the Sly EoS. 

As a final note on the prior construction described above, although we label them model-agnostic and model-informed priors, in relation to previous analyses (e.g. ref.~\cite{Landry:2019}), our draws still closely follow the class of theoretical models they were trained upon. We refer to informed training in the sense of grouping similar datasets together, and then creating multiple GMMs whose combined draws constitute the prior. In contrast, in the agnostic training, all datasets are treated equally and taken together to build a single GMM. Similarity of datasets is given by particle species treated by the theoretical models. No assumptions are added about resulting EoS properties (such as imposing that they support a given NS mass), nor is any artificial noise injected after the GMM is drawn. Combining all this with optimizing the cross validation likelihood, we construct a set of realistic equations of state, even in the model agnostic prior, as opposed to a prior containing maximum model ``variability''. Furthermore, in our case, it is the large sample of training EoSs that attempts to mitigate the level of information gained by choosing a specific model. Within this non-parametric prior construction scheme, the ``variability'' of draws should come from hyper-prior and even kernel selection more than from the exact EoSs chosen if one aims to construct a space of ``all possibilities'', and thus a true model-agnostic prior. This is why we present the second prior, rather than just the first, to illustrate the available capacity of the data to distinguish between the different options of NS EoSs.

\section{Methods} 
\label{Sec:methods}

The literature on constraining the equation of state of cold nuclear matter with multi-messenger observations is quite broad, see e.g. refs.~\cite{Agathos:2015uaa,2015PhRvD..91d3002L,2016EPJA...52...69A,2017ApJ...850L..19M,2018PhRvL.120q2703A,2018PhRvL.120z1103M,2018ApJ...852L..29R,2018ApJ...852L..25R,2018MNRAS.478.1093R,2018PhRvC..98d5804T,2019MNRAS.485.5363G,2019ApJ...876L..31K,Landry:2019,2019PhRvD.100b3015S,2020NatAs...4..625C,2020EPJST.229.3663C,2020Sci...370.1450D,Landry:2020vaw,Raaijmakers:2019dks}. It is common, to place hard constraints on certain macroscopic properties of the EoS, e.g., when implementing previous high mass pulsar observations. For example, ref.~\cite{LIGOScientific:2018cki}, allowed only draws from a prior on the EoS parameters that support a maximum mass consistent with, say PSR J0348+0432~\cite{Antoniadis:2013pzd}, with $M=2.01\pm0.04\,M_\odot$. This does not utilize the full information present in that observation, i.e., the whole posterior distribution. Moreover,  as noted in ref.~\cite{Miller:2019nzo}, this makes the probability of an EoS supporting a maximum allowed NS mass of say $M_\text{max}=2.07\,M_\odot$ equal to that of supporting $M_\text{max}=2.7\,M_\odot$, both having a probability of exactly one. This obviously cannot be the case, as ideally, the observation of PSR J0348+0432 would significantly favor the former EoS as opposed to the latter. 
Furthermore, there is now a growing body of evidence favoring a bimodal mass distribution of neutron stars \cite{Alsing:2017bbc,Fan:2023spm}. This indicates that setting an a priori cut-off either in the sense of removing EoSs that do not support a given mass or even treating the observation of a given mass as being equal proof for or against two very different EoSs, phenomenologically speaking, is perhaps not an optimal strategy. The use of such strict bounds also disallows the integration of multiple datasets, such as using both the observation of PSR J0348+0432 and PSR J0740+6620~\cite{Miller:2021qha} with $M=2.14^{+0.10}_{-0.09}\,M_\odot$. In order to alleviate this, along with the comments on EoS parametrization model bias, our method treats the draws of a GMM prior as a categorical distribution, which is then sequentially updated with different data likelihoods, based on full posteriors, in a Bayesian scheme. We shall now describe in more detail this methodology.

The aim is to combine four datasets $\bm{d}_\text{GW}$, $\bm{d}_\text{X-RAY}$, $\bm{d}_\text{M}$, $\bm{d}_\text{pQCD}$ into a constraint on the cold, catalyzed, dense matter EoS, using a Bayesian procedure, so the posterior probability for a single EoS $\alpha$, given the available data $\bm{d}=\{\bm{d}_\text{GW}, \bm{d}_\text{X-RAY}, \bm{d}_\text{M}, \bm{d}_\text{pQCD}\}$, is
\begin{equation}
    \mathcal{P}(\alpha\vert \bm{d}) = \frac{\mathcal{L}(\bm{d} \vert \alpha) \varphi (\alpha)}{\mathcal{Z}(\bm{d})}=\frac{\varphi(\alpha) \prod_i \mathcal{L}(\bm{d}_i \vert \alpha) }{\mathcal{Z}(\bm{d})},
\end{equation}
with $i$ enumerating the datasets, and the form of the likelihood $\mathcal{L}(d_i \vert \alpha)$ depends on the specific dataset, see Section~\ref{Sec:data}. $\varphi(\alpha)$ represents the probability of an EoS $\alpha$ prior to analysis, while the evidence $\mathcal{Z}(\mathbf{d})$ acts as a normalization factor.

Previous work using Gaussian process EoS priors~\cite{Landry:2019,Essick:2019ldf,Landry:2020vaw, Essick:2023fso} all used Monte Carlo (MC) integration to estimate the posterior, i.e., randomly picking a given draw from the prior (along with draws from mass priors and other parameters, depending on likelihood form), computing the likelihood, and repeating this process until a sufficiently large number of samples have been collected. This is the standard approach when performing Bayesian inference with a Gaussian process prior, and it is also one of the reasons why GPs are not more widely adopted in non-parametric Bayesian inference schemes (as opposed to other machine learning methods such as deep neural networks, e.g. refs.~\cite{Fujimoto:2017cdo,Fujimoto:2019hxv,Fujimoto:2021zas,Fujimoto:2024cyv}). The freedom to better describe EoSs with non-parametric models is always coupled with the additional computational cost this increased complexity entails. This cost is partly due to the training required to construct the GMM in the first place, and it scales up with the number of EoSs included, and partly due to the way the draws themselves have to be treated in direct EoS inference schemes, namely MC integration. Thus, the problem of the standard approach is computational efficiency of MC integration and data stream inflexibility. Specifically, the standard treament of EoS draws from a non-parametric model, is to compute a likelihood of the form, $\mathcal{L}(\bm{d}_i \vert \alpha)$, where the likelihood is written as a function of the EoS itself only, as opposed to a set of parameters including the EoS. This means that one is somewhat forced to construct a Kernel Density Estimator (KDE) of some data posterior (say mass-radius posterior of NICER's PSR J0740+6620 measurement) and compute weights directly. Therefore, when utilizing non-parametric priors in previous GW EoS inference scenarios, it was not possible to simultaneously infer the EoS (drawn from a non-parameteric prior), and other source parameters characteristic of the data stream (for example in the case of GWs of BNS coalescence, component masses, spins etc).  This is in contrast to parametric methods, where in virtue of having a finite set of parameters, one can utilize raw data to constrain those parameters, as was done  e.g. in  ref.~\cite{LIGOScientific:2018cki}. In other words, parametric models are very easy to integrate in existing data analysis pipelines, e.g.  \texttt{Bilby} already has support for replacing the traditional inference basis of tidal deformabilities with four parameters for the piecewise polytrope or spectral decomposition models. The method we will describe next allows simultaneous sampling of the parameters of neutron star mergers and the equation of state, without a parametric EoS construction. 

In order to avoid these common issues, we introduce a novel and flexible treatment of GP prior draws that is computationally efficient (in virtue of being able to use both Markov Chain Monte Carlo (MCMC) and nested sampling algorithms) and is easily integrated into raw data analyses (such as parameter estimation of GW170817, which is showcased in Section~\ref{Sec:data}). We first construct a categorical distribution (Generalized Bernoulli distribution) over $n$ draws from a Gaussian process (or Gaussian mixture model in general), which ought to constitute a large representative sample of the underlying distribution. A categorical distribution is a discrete probability distribution that describes the outcomes of a finite set, with probability mass function for $n$ outcomes:
\begin{equation}
    P(X=i)=\pi_i,\quad i=1,2,...,n,
\end{equation}
where $X$ is a random variable representing the outcome, $\pi_i$ is the probability of the outcome $i$, and $\sum^n_{i=1} \pi_i=1$. The cumulative distribution function is then
\begin{equation}
    F(X\leq i)=\sum^i_{j=1}\pi_j,\quad i=1,2,...,n .
\end{equation}

In our case, each outcome is a given EoS draw from a GMM, and as such, from the point of view of the nested sampler/MCMC algorithm we have parametrized a non-parametric distribution. A single parameter, EoS category (from now on referred to as EoS index), is sampled, by merely sampling in the uniformly distributed $[0,1]$ interval and finding the smallest integer $j$ such that our random number $u\leq \sum^j_{i=1}\pi_i$, with the integer $j$ being the relevant EoS. This allows us to preserve the full EoS information (no parameterization), but use our GMM draws in standard, and efficient, parameter estimation schemes, as opposed to having to do MC integration on likelihood weights. Within any parameter estimation pipeline, the sampler is often exposed to conversions of the actual relevant likelihood parameter, for convergence purposes, and this is the same principle, but without loss of structure of the underlying object of inference, the EoS. Prior to exposure to data, our categorical distribution is equally weighted, and after a Bayesian updating step, we will have a posterior over the EoS index that will be an unequal-weighted categorical distribution. 

Another advantage of using a discrete distribution is that, by definition, the order of the categories does not matter. Therefore, we can arrange the prior in any way, and even re-arrange it after a given Bayesian step, without affecting the resulting posterior, beyond merely reordering it. The obvious advantage of this is that we can sort the prior based on whatever parameter is most relevant for the likelihood. For example, the gravitational waveform data from BNS mergers is most sensitive to the underlying NS EoS through the tidal deformability parameter at a given mass. Thus, when incorporating the GW170817 event we can sort our GMM draws by the predicted tidal deformability at $1.4 M_{\odot}$, a unique value for any given $\epsilon(p)$ relation.  The specific choice of mass at which to compute $\Lambda$, only barely affects the sorting, within the GW170817 mass range, and thus a negligible impact on the evidence estimation speed. In effect, we are now sampling tidal deformabilities when considering this dataset, which is naturally much more efficient than MC integration of draws and can easily be integrated into existing gravitational wave parameter estimation pipelines. In other words, the nested sampler provides likelihood values for each sample as if it was directly sampling from a traditional tidal deformability prior, so we do not suffer from potentially extreme likelihood surfaces. A similar argument can be made in the context of the NICER observations, where instead of mass and compactness as sampled parameters, we have mass and EoS index (from which we can compute radius, then compactness to pass to the likelihood), to directly constrain the EoS.

In summary, our analysis consists of the following steps:
\begin{enumerate}
    \item Construction of a GMM of GPs from 75 tabulated theory EoSs, hadronic, hyperonic and quark models.
    \item Drawing 10000 samples from the GMM and creating a categorical distribution over them as a prior stand-in.
    \item Sorting the prior by expected tidal deformability at $1.4\,M_\odot$, as the EoS-dependent variation on $\Lambda$ within the $\sim1.25\,M_\odot-1.5\,M_\odot$ event range is negligible (the exact choice of the mass would not substantially change the order). 
    \item Running a parameter inference scheme of the BNS event GW170817 with one out of the 16 sampled parameters as the EoS index, instead of individual tidal deformabilities. 
    \item Using the resulting unequal-weighted categorical distribution posterior over the EoS index as a prior in the next Bayesian step and re-ordering it by maximum supported NS mass.
    \item Kernel density estimators are built for each of the mass-radius posteriors of two NICER observations and Gaussians for the 105 NS mass measurements.
    \item We estimate the evidence for this Bayesian update with nested sampling.
    \item The posterior inferred is resorted again, now by predicted pressure at six times saturation density.
    \item Using the likelihood and method of ref~\cite{Komoltsev:2023zor} we further narrow the posterior through a Bayesian updating step including results from pQCD calculations.
\end{enumerate}

\section{Multi-Messenger observations and pQCD calculations} 
\label{Sec:data}

In this section, we describe our analysis of the GW170817 event using our methodology in a modified version of the \texttt{Bilby} gravitational wave (GW) inference pipeline \cite{Ashton:2018jfp,2019ApJS..241...27A}. Our reason for re-running the GW analysis, as opposed to just using posteriors as we do with other NS observations, is two-fold. Firstly, to showcase the ease of running a parameter inference scheme on raw data with complex and large parameter spaces, with existing pipelines, even when using non-parametric priors. Second, the best direct constraints on the NS EoS from the GW170817 event, that utilize the gravitational waveform itself in the analysis, are done using parametric schemes. Moreover, even utilizing the most unassuming posterior, the original event identification \cite{LIGOScientific:2017vwq}, would mean a posterior over the 5PN and 6PN order normalized tidal deformabilities, with only the former being well constrained:
 \begin{equation} 
    \Tilde{\Lambda}=\frac{16}{13}\frac{(M_1+12M_2)M_1^4\Lambda_1+(M_2+12M_1)M_2^4\Lambda_2}{(M_1+M_2)^5},
\end{equation}
with $\Lambda_{1,2}$ and $M_{1,2}$ the component tidal deformabilities and masses respectively.
It is clear that in order to break the degeneracy and constrain $\Lambda_1$ and $\Lambda_2$ requires further assumptions, with universal relations being the most common tactic~\cite{Yagi:2015pkc,Yagi:2016bkt}. 

The posteriors obtained from the GW170817 event, will then be used as a prior for the next data integration. We also outline the construction of the likelihood for two NICER observations, along with 105 NS mass observations, where the posteriors are utilized from the respective event detection analyses. Lastly, we include recent pQCD calculations~\cite{Gorda:2022lsk,Komoltsev:2023zor,Gorda:2023usm}.

\subsection{Gravitational waves}

\begin{table*}
\centering
\caption{Summary of the priors used for parameter estimation of the GW170817 BNS event. The detector-dependent ``extrinsic'' parameters (distance, sky localization, angle between line of sight (LOS) and orbital angular momentum (OAM), polarization angle, phase at and time of coalescence) are listed in the first box, while the rest are the system ``intrinsic'' parameters (component spins, masses and EoS).  The chirp mass and time of coalescence are based on the results from ref.~\cite{LIGOScientific:2017vwq}, while the rest of the priors are taken/inspired from ref.~\cite{LIGOScientific:2018cki}.  The EoS index for the  the model-informed prior constructed analysis from 3 GMMs, trained on hadronic, hyperonic and quark EoSs respectively, has 10000 draws each.  } \label{tab:Priors}
\begin{tabular}{|l|c |c|}
\hline
\multicolumn{3}{|c|}{GW170817}\\
\hline
 & Parameter & Prior Type\\
 \hline
Comoving distance & $d_L$ [Mpc] & Uniform $[10,100]$\\
Right ascension & $\alpha$ [h m s] & Gaussian $[\mu=13^h09^m48.085^s,\sigma=0.018^h]$\\
Declination & $\delta$ [degrees] & Gaussian $[\mu=-\ang{23;22;53.343},\sigma=\ang{0.218}]$\\
Angle(LOS, OAM) & $\theta_{JN}$ [rad] & Sine $[0,\pi]$ \\
Polarization & $\psi$ [rad] & Uniform $0,\pi]$\\
Phase at coalescence & $\phi_c$ [rad] & Uniform $[0,2\pi]$\\
Time of coalescence & $t_c$ [s] & Gaussian $[\mu=1187008882.43,\,\sigma=0.002]$\\
\hline
Spin magnitudes & $a_1,a_2$ & Uniform $[0,0.05]$ \\
Tilt angles & $\theta_1,\theta_2$ [rad] & Sine $[0,\pi]$\\
Azimuthal separation & $\delta\phi$ [rad] & Uniform $[0,2\pi]$\\
Cone of precession & $\phi_{JL}$ [rad] & Uniform $[0,2\pi]$\\

\hline
Chirp mass & $\mathcal{M}_c$ [$M_\odot$] & Uniform in components $[1.197,1.198]$\\
Mass ratio & $q$  & Uniform in components $[0.125,1.0]$\\
\hline
\multicolumn{3}{|c|}{EoS Index}\\
\hline
Model agnostic& $e$ & Equal-Weighted Categorical $[\text{categories}=10000]$\\
\hline
Model informed& $e$  & Equal-Weighted Categorical $[\text{categories}=30000]$\\
\hline
\end{tabular}
\end{table*}

Gravitational wave observations of binary neutron star mergers can constrain tidal deformabilities and NS masses, which, in turn, constrain the EoS. Given an equation of state,  using the TOV equations and the tidal Love number \cite{hinderer2008tidal,hinderer2010tidal}, 
the macroscopic parameters: stellar mass, radius and tidal deformability in the inspiral phase are inferred.

Perhaps the most significant neutron star observation to date is the August 17, 2017, detection of a gravitational wave transient signal, GW170817~\cite{LIGOScientific:2017vwq}. Both the Advanced LIGO detectors at Hanford, Washington (H1) and Livingston, Louisiana (L1), and the Advanced Virgo (V1) detector in Pisa, Italy were in operation at the time, and contributed to constraining  the nature of the source and its properties. The gravitational wave signal was followed by an electromagnetic (EM) counterpart \cite{LIGOScientific:2017ync, LIGOScientific:2017zic}, which solidifies the cause of this signal being a merging BNS system. The short gamma-ray burst was independently observed by the Fermi Gamma-ray Burst Monitor and the Anti-Coincidence Shield for the Spectrometer for the International Gamma-Ray Astrophysics Laboratory~\cite{Goldstein:2017mmi,2017A&A...603A..46S}. Although a neutron star-black hole (BH) system is not completely ruled out, the inferred masses and spins are characteristic of the known NS population \cite{Wysocki:2020myz}. The signal-to-noise ratio (SNR) of the event is $\sim 3$ higher than that of  the other most likely BNS GW event, GW190425~\cite{LIGOScientific:2020aai}, which makes it ideal for obtaining quality constraints on the underlying cold, high-density nuclear matter EoS. Note that the nature of the source system's components of the GW190425 signal is much more uncertain. This source uncertainty is a problem mostly unique to GW observations, since without an EM counterpart (which is not necessarily a given), distinguishing between NS-NS, NS-BH and BH-BH coalescence events can be potentially problematic, as is the case of  GW190425. In this sense, it is worth clarifying, that we assume the source of GW170817 is a BNS in order to perform our analysis, an assumption that cannot necessarily be easily extended to other potential BNS candidates such as GW190425, GW190814~\cite{LIGOScientific:2020zkf}, GW191219~\cite{KAGRA:2021vkt}, GW200105 or GW200115~\cite{LIGOScientific:2021qlt}, and GW230529~\cite{LIGOScientific:2024elc}, which is why these are excluded from this work.

For the GW analysis, we use $256\s$ of the GW170817 merger event data from H1, L1 and V1 detectors, centered such that the segment ends $2\s$ after the estimated merger time, and a Power Spectral Density (PSD)  of $1024\s$ of data. This data is taken from the second version of the official LIGO data release, and corresponds to the cleaned strain \footnote{There is a glitch in the L1 data segment which was modeled and subtracted, along with other known noise sources, in the second data release. This is also the version used in ref.~\cite{LIGOScientific:2018cki}.}. More information about the noise modeling can be found in ref.~\footnote{\url{https://www.gw-openscience.org/events/GW170817/}}. We also analyze the full frequency range of the data release, i.e., $20-16384\Hz$. Although the signal enters the detector's sensitivity range less than $100\s$ before the merger and the calibration model of the detectors covers at best the $20-4096\Hz$ range.  This data was selected to allow for higher resolution sampling. Marginalization over the detector's calibration uncertainty has a limited impact on the final result, as the difference between distinct calibration models are much smaller than the inherent statistical uncertainty of the data. For an in-depth comparison of physically motivated calibration models and their effects on the source properties of all events in GWTC-1 (first LIGO observing run) see  ref.~\cite{Payne:2020myg}.

For the waveform approximant, we have chosen the most accurate phenomenological model available in \texttt{lalsuite}~\cite{lalsuite,swiglal} to date (April 2023), which allows a tidal-phasing extension, \emph{IMRPhenomPv2} \cite{Hannam:2013oca,Husa:2015iqa} as the base BH approximant. The tidal addition from numerical relativity tuning is \emph{NRTidalv2} \cite{Dietrich:2019kaq}. This model allows non-aligned spins, as well as precession effect for the dominant $(l, \lvert m \rvert)=(2,2)$ mode. Although effective one body (EOB) approximants are more accurate in the later inspiral phase, they are drastically computationally slower (up to CPU hours per orbit evolution for \emph{SEOB}~\cite{Taracchini:2013rva} as opposed to fractions of a CPU second for \emph{IMRPhenomPv2}) and have a direct dependency on the signal duration. The purpose of the waveform approximant is to model the BNS dynamics and thus the effect on the GW generation by providing a waveform as a function of the source parameters. Thus, given a gravitational wave data strain $d_\text{GW}$, a waveform template $\mu_{\text{IMR}}$ as a function of $\theta$, the ``search parameters'' in Table~\ref{tab:Priors},  and the detector noise $s$, we have the typically assumed Gaussian-noise likelihood~\cite{Thrane:2018qnx}:
\begin{equation}
    \mathcal{L}_\text{GW}(d_\text{GW};\theta)=\frac{1}{2\pi s ^2}\exp\left(-\frac{| d_\text{GW}-\mu_{\text{IMR}}(\theta)| ^2}{2s ^2}\right),
\end{equation}
where the data strain is complex valued. Note the lack of a square root in the normalization. In practice however, we actually utilize the relative binning likelihood introduced in ref.~\cite{Zackay:2018qdy}, see Appendix~\ref{App:gw170817}.

Our method is integrated in the \texttt{Bilby} pipeline by modifying the conversion function of the likelihood, which for the waveform approximant used, computes the gravitational waveform from a tuple of masses and tidal deformabilities for the two NSs involved. In previous parametrized analyses (e.g. ref.~\cite{LIGOScientific:2018cki}), the two tidal deformabilities were replaced with a set of EoS parameters from which the deformabilities were inferred, effectively sampling directly in the EoS space as opposed to macroscopic observables. Similarly, we utilize one parameter, the EoS index, together with a pair of masses, to compute the likelihood. The priors utilized for this stage can be found in Table~\ref{tab:Priors}.

\subsection{X-ray and radio observations}

Some isolated NSs exhibit pulse X-ray waveform emission profiles, which are thought to be produced by hot spots on their surface \cite{Landry:2020vaw}. These small regions are created by cascades of particles from pair production in magnetosphere plasma gaps. The rotational period of the star determines the light curve of the thermal X-rays in these heated regions, with the shape correlated with the stellar compactness $C$, as the period of visibility increases at the expense of the totality of the eclipses. Once the compactness is determined, the radius can be measured as relativistic Doppler and aberration effects in the light curve scale with the size of the object and its rotation rate.

Two independent analyses were performed on the pulse profile of a specific pulsar, J0030+0451 \cite{Miller:2019cac,Riley:2019yda}. Both studies concluded that the pulsar's waveform suggests the emission comes from two or three non-circular hot spots. Multiple hot spot models were considered, and all produced consistent results for the mass and radius of the pulsar. Importantly, despite differences in the hot spot geometry assumed in different models, the constraints on the EoS and NS observables appeared consistent within statistical error. This implies that such systematic differences are minor enough not to substantially affect the derived EoS information. Similarly, we have the NICER and X-ray Multi-Mirror (XMM)-Newton X-ray
observations of PSR J0740+6620 \cite{Miller:2021qha}. For the NICER+XMM-Newton data, we utilize the full mass radius posterior provided in refs.~\cite{Miller:2019cac} and \cite{Miller:2021qha} when computing our likelihood~\footnote{ 
The posteriors for PSR J0030+0451 and PSR J0740+6620  are available at \url{https://zenodo.org/record/3473466} and \url{https://zenodo.org/record/4670689}, respectively.}. 
The mass-radius likelihood is obtained by integrating the full mass-radius curve predicted by a given EoS, so for an observation $i$, we have~\cite{Miller:2019nzo}:
\begin{equation} 
\label{eq:LikeNICER}
    \mathcal{L}_{R,i}(\bm{\alpha})=\int dM \pi_i(M) L_i(M, R(M, \bm{\alpha})),
\end{equation}
with $\pi_i(M)$ being the NS observation's mass prior (uniform). The likelihood of a given star $i$, to have a mass $M$, and radius $R$, is $L_i(M,R)$, so a KDE constructed from the mass-radius posterior samples, $L_i(M, R(M, \bm{\alpha}))$ is evaluated.
It is worth remarking that we have assumed a uniform mass distribution, lacking a better understanding of NS formation mechanisms and population statistics. Properly accounting for the latter can bias the inferred EoS. Specifically, there is growing reason to believe  that the NS population distribution is multi-modal, see e.g. ref.~\cite{Wysocki:2020myz}. Similarly, the presence of phase transitions in some but not other NSs, leads to vastly different radii, all contributing to biasing EoS inference if not accounted for. 

Next, there are radio observations of massive pulsars, their main role being increasing the maximum allowed mass of a given EoS. Since a candidate EoS being stiffer or softer determines the maximum mass a neutron star can support, any observation of the most massive pulsars can directly rule out the softest  EoS models. The pulsars we list in Appendix \ref{App:NSobs} are a sample of 105 pulsars with individual mass measurements (which also includes low massive NSs). They are a mix of NS-NS/White Dwarf/Main sequence star systems, millisecond pulsars, Black Widow and Redback, isolated NSs and High/Low mass X-ray binaries. We include this data by taking Gaussians around the values quoted in Table~\ref{tab:NScatalog}. The mass likelihood is obtained by simply integrating over the mass posterior up to the EoS's  maximum allowed mass, so for a given mass posterior $P_i(M)$ of a particular NS we have~\cite{Miller:2019nzo}:

\begin{equation} \label{eq:LikeMass}
    \mathcal{L}_{M,i}(\bm{\alpha})=\int_0^{M_\text{max}(\bm{\alpha})} P_i(M) dM.
\end{equation}
For a given NS mass observation with quoted mean $\mu_i$ and standard deviation $\sigma_i$ (see Table~\ref{tab:NScatalog} for the specific values), a Gaussian distribution for the mass posterior is constructed, $P_i(M)=\mathcal{N}(\mu_i(M),\sigma_i(M))$. Furthermore, by having the termination condition of the integral corresponding to the maximum supported mass of an EoS, we implicitly assume that NS mass population cutoff corresponds to said mass, as opposed to a lower value, see~\cite{Golomb:2024lds} for a discussion on this point.

\subsection{Perturbative QCD calculations}

Nuclear matter in the core of stable NSs can reach, depending on the particular EoS choice, densities of $\sim5-8 \, n_\text{sat}$, where $n_\text{sat}$ is the nuclear saturation number density. In this regime QCD, the theory of strong interactions is non-perturbative. However, at high density $\sim20-40\,n_\text{sat}$, hence high energy, it becomes perturbative. The connection between these two regimes requires extrapolating either of these calculations \cite{Kurkela:2014vha,Komoltsev:2021jzg}.

Very recently, a potentially model-agnostic way of including the pQCD input, by exploring the connection between the low-density NS regime and the high-density perturbative regime, has been suggested \cite{Gorda:2022jvk, Komoltsev:2023zor,PhysRevLett.127.162003}. The likelihood is constructed using the requirement that the triplet formed by pressure, baryon number density and chemical potential, $\bm{\beta}\equiv\{p,n,\mu\}$, can reach the QCD values at high densities with a stable and causal EoS. Each EoS, will predict a unique pressure and number density at a given baryon chemical potential, so the triplet needs to stably and causally connect the low density/EoS termination regime, $n_L\equiv n_{term}$, to $n_H\equiv40\,n_\text{sat}$,  which will be the integration interval of the likelihood \cite{Komoltsev:2023zor} 
\begin{equation} 
\label{eq:LikeQCD}
    \mathcal{L}_\text{pQCD}(\bm{\alpha})=\int d (\log \bm{X}) w(\log\bm{X})  \bm{1}_{[0,1]}(I_{\text{pQCD}}(\bm{X}, \bm{\alpha})),
\end{equation}
where, the pQCD tension index, $I_{\text{pQCD}}(\bm{X}, \bm{\alpha})~\equiv(\Delta p -\Delta p_{min})/(\Delta p_{max}-\Delta p_{min})\in [0,1]$, is used to quantify the agreement between a given EoS and the high-density regime. $\bm{1}_{[0,1]}(I_{\text{pQCD}})$ assigns a zero Bayesian weight to an EoS for whom $\bm{\beta}$ violates $\Delta p=p_H-p_L\in [\Delta p_{min},\Delta p_{max}]$, and $\bm{X}\equiv 3\overline{\lambda}/(2\mu_H)$ is the (dimensionless) renormalization scale. The minimum and maximum pressure intervals allowed correspond to EoSs which will have one segment of $c_s=1$ and a subsequent phase transition or vice versa, respectively.

In practice, the integral of Eq.~\ref{eq:LikeQCD} is evaluated with Monte-Carlo integration by randomly picking values from the distribution $w(\text{log}\,X)=\bm{1}_{[\text{ln}(1/2),\text{ln}(2)]}(\text{log}X)$ and counting the frequency that the termination triplet, satisfies the above condition, as in ref.~\cite{Cacciari:2011ze}. For a complete explanation of this likelihood, see ref.~\cite{Gorda:2022jvk}. We utilize the ``prior'' marginalized likelihood, introduced in ref.~\cite{Komoltsev:2023zor}, without the addition of the speed of sound constraint below $40\,\nsat$, and choose the mean of our number density prior distribution as the EoS termination point, $n_\text{term}$~\footnote{A python implementation of this likelihood construction can be found at \url{https://zenodo.org/records/10592568}}.

\section{Results}
\label{Sec:results}

\begin{figure}
    \centering
    \includegraphics[width=\columnwidth]{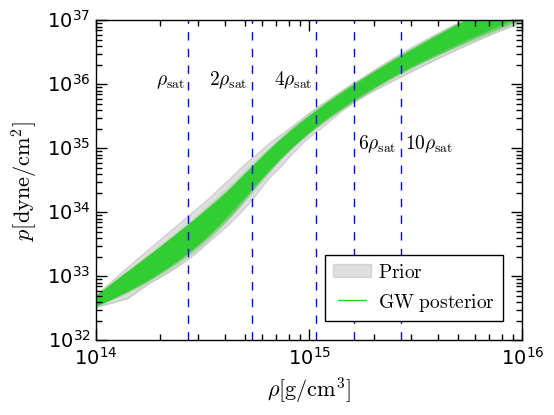}
    \caption{$95\%$ confidence interval of the posterior EoS distribution after the GW170817 event analysis. For reference, the model agnostic prior is depicted in gray. Vertical dashed blue lines represent once, twice, four, six and ten times the nuclear saturation density, $\rho_{\text{sat}}=2.7\times10^{14}\g/\cm^3$.}
    \label{fig:agn_result_gw}
\end{figure}

We start by discussing the parameter inference of the GW170817 waveform, within our modified scheme, which resulted in standard estimates of the source properties. Assuming a BNS event and the same EoS for both NSs, we constrain the component masses, $M_1 = 1.47^{+0.13}_{-0.10}M_\odot$ and $M_2 = 1.274^{+0.091}_{-0.10}M_\odot$, and tidal deformabilities, $\Lambda_1 = 339^{+200}_{-200}$ and $\Lambda_2 = 791^{+500}_{-400}$, within a $95\%$ confidence interval, see Appendix \ref{App:gw170817} for full posteriors. In terms of providing direct equation of state constraints, the relatively low masses involved, only manage to narrow the posterior slightly around twice the nuclear saturation range, as was previously found (see e.g. refs.~\cite{LIGOScientific:2018cki,Landry:2020vaw}). Figure~\ref{fig:agn_result_gw} illustrates in the standard baryon density versus pressure plane how the GW posterior has little to no impact on the very high or very low density regimes, the former not being realistically achievable for a $\sim 1.4M_\odot$ NS. 

\begin{figure}
    \centering
    \includegraphics[width=\linewidth]{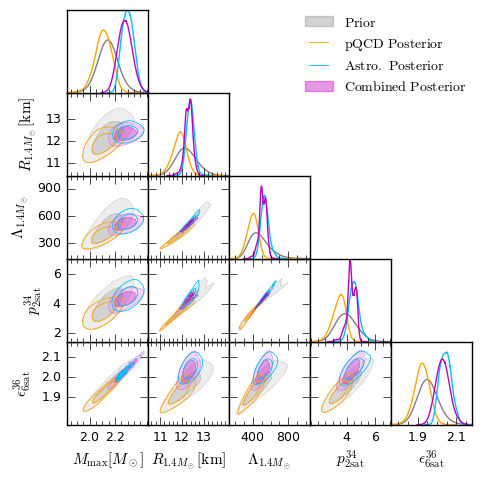}
    \caption{Corner plot of macroscopic (maximum supported mass, predicted radius, and tidal deformability of a $1.4M_\odot$ NS) and microscopic (pressure at twice nuclear saturation density, and energy density at six times nuclear saturation density) EoS properties. The prior is shown in gray,   posteriors after inclusion of either astrophysical observations or pQCD  are depicted in light blue and  orange, respectively, and the combined posterior (astrophysical and pQCD) is shown magenta. 
    Pressure and energy density are given in units of $10^{34}$ and $10^{36}{\rm dyne/cm^2}$, respectively.}
    \label{fig:corelations}
\end{figure}

For this analysis, the EoS index prior was sorted by the expected tidal deformability of a $1.4M_\odot$ NS, due to this being the most relevant prediction of a given EoS for such event. There is a strong correlation (Pearson Coefficient (PC) $\sim 0.98$), at least within our EoS prior, between tidal deformability of a $1.4M_\odot$ NS and pressure at twice nuclear saturation density. 

\begin{figure}
    \centering
    \includegraphics[width=\columnwidth]{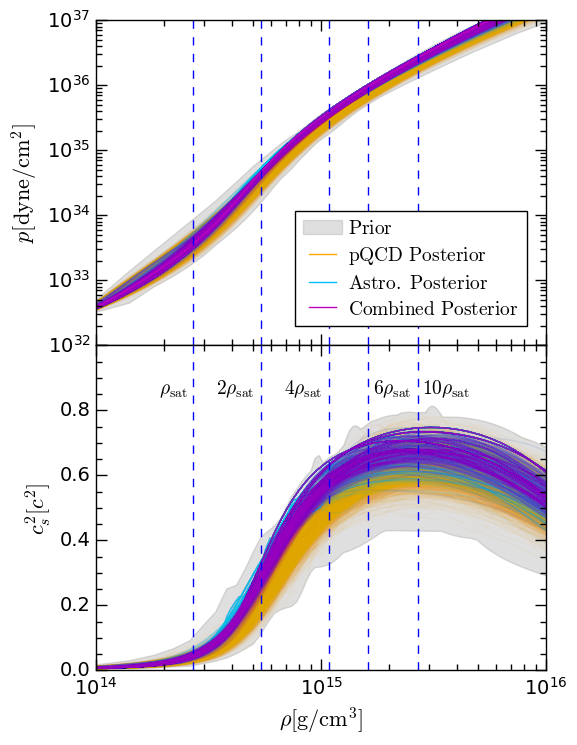}
    \caption{$95\%$  confidence interval of the posterior EoS distribution after including  either astrophysical observations (light blue) or pQCD (orange), and the combined posterior (observations and pQCD) (magenta) in baryonic density versus pressure (left), speed of sound as fraction of speed of light (right) representation. For reference, the model agnostic prior is shown as a gray background. Vertical dashed blue lines represent once, twice, four, six and ten times the nuclear saturation density, $\rho_{\text{sat}}=2.7\times10^{14}\g/\cm^3$.}
    \label{fig:agn_result}
\end{figure}

Next, we show the effect of including all astrophysical observations, namely X-rays (NICER), radio and GWs, as well as pQCD calculations. This is illustrated in  Figure \ref{fig:corelations}. 
The samples shown are calculated for any given EoS in a posterior/prior distribution. Using astrophysical data alone, we constrain the pressure at twice nuclear saturation to  $p_{2\rho_\text{sat}}^\text{Astro} = \,4.5^{+0.8}_{-0.7}\,\times 10^{34}\,\text{dyne/cm}^{2}$, and adding the pQCD result we find $p_{2\rho_\text{sat}}^\text{Astro+pQCD} = \,4.3^{+0.6}_{-0.6}\,\times 10^{34}\,\text{dyne/cm}^{2}$.

\begin{figure}
    \centering
    \includegraphics[width=\columnwidth]{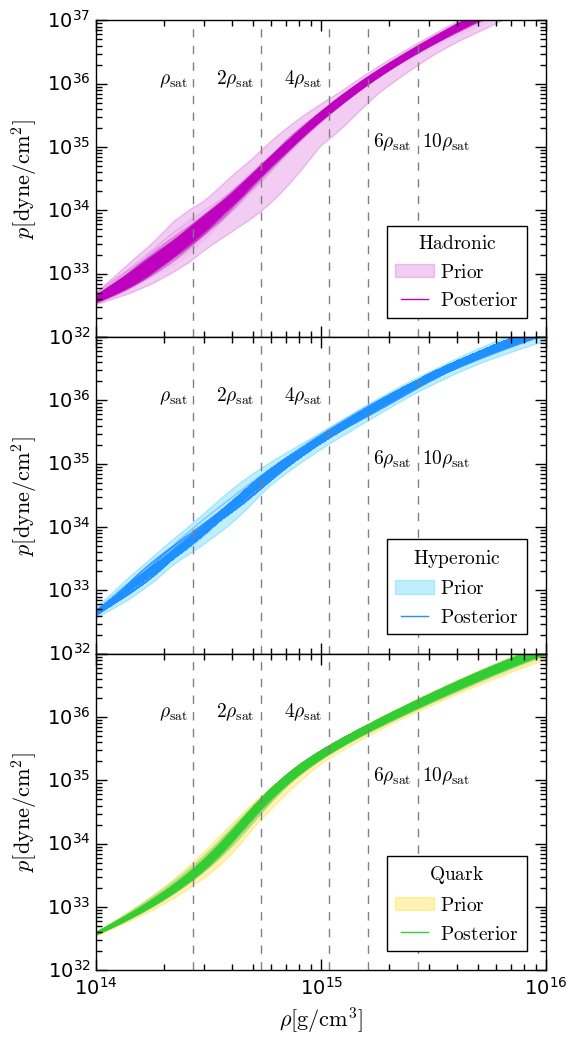}
    \caption{$95\%$ confidence interval of the posterior EoS distribution after inclusion of all datasets (astrophysical + pQCD) for our separate prior, composed of the hadronic GMM (magenta), hyperonic (light blue) and quark (yellow). For reference, vertical dashed gray lines represent once, twice, four, six and ten times the nuclear saturation density, $\rho_{\text{sat}}=2.7\times10^{14}\g/\cm^3$.}
    \label{fig:Sep_post}
\end{figure}

Another correlation (PC $\sim 0.95)$, between maximum supported NS mass and energy density at $6\,\rho_\text{sat}$, along with the likelihood construction, Eqs.~\ref{eq:LikeNICER} and \ref{eq:LikeMass}, provides the strong constraints on the high density regime observed in Figure~\ref{fig:agn_result} for the astrophysical posterior (light blue). The maximum supported NS mass in the astrophysical posterior is constrained to, $M_\text{max}^\text{Astro} = 2.300^{+0.090}_{-0.087}\,M_\odot$, while including pQCD data, $M_\text{max}^\text{Astro+pQCD} = 2.28^{+0.11}_{-0.11}\,M_\odot$. Both the $p(\rho)$ and $c_s(\rho)$ views of the EoS are driven to higher pressures/speeds of sound in the $\sim6 - 10\, \rho_\text{sat}$ regime by the astrophysical data. Again, this is to be expected, as the larger the NS mass the higher the core energy density (and pressure, see Appendix~\ref{App:Agn Posterior} for full posteriors). From astrophysical data, the pressure at six times nuclear saturation density is constrained to, $p_{6\rho_\text{sat}}^\text{Astro} = \,9.0^{+1.0}_{-1.0}\,\times 10^{35}\,\text{dyne/cm}^{2}$, while including pQCD calculations, barely affects the posterior, as the pQCD posterior closely follows the prior. In contrast, the low mass NS observations, see Appendix~\ref{App:NSobs}, have no effect on our prior, due to the likelihood construction, namely the integral of Eq.~\ref{eq:LikeMass} adds up to unity for a low mass, narrow Gaussian posterior. Echoing the earlier discussion on likelihood construction, this is perhaps a reason why a penalty term on EoSs that support a much higher mass than the most massive pulsar observed ought to be included in the likelihood. That is, until a more complete NS population statistic can be constructed, which naturally would require a better understanding of NS formation as well as parametrized modeling of the accretion/spin-up phase. For an explanation of why incomplete NS population statistics can bias all such analyses, see ref.~\cite{Miller:2019nzo,Wysocki:2020myz}.

\begin{figure}
    \centering
    \includegraphics[width=\linewidth]{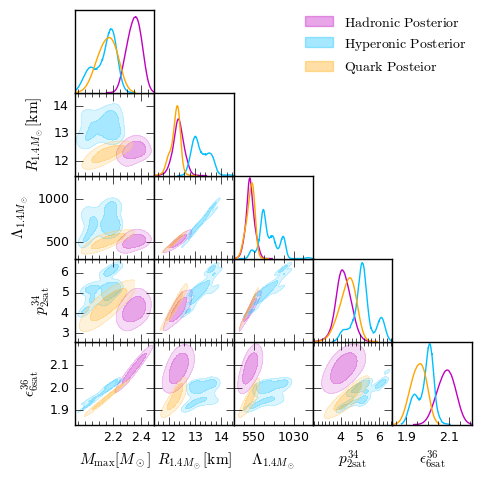}
    \caption{Corner plot of the posteriors of the same macroscopic and microscopic properties as in Fig.~\ref{fig:corelations}, but for the model informed prior. Each combined  posterior, after inclusion of astrophysical observations and pQCD calculations for the hadronic (magenta), hyperonic (light blue) and quark (orange) is shown.}
    \label{fig:corelations_informed}
\end{figure}

In terms of the separate Gaussian mixture model prior, separately trained on hadronic, hyperonic and quark models respectively, Figure~\ref{fig:Sep_post} illustrates in the pressure-density plane the combined posterior (all observations and pQCD). We immediately note that the hadronic prior is the best constrained, while the hyperonic/quark posteriors are prior-dominated. This is due to the preference of astrophysical data for heavy supported masses, and the narrow prior range. On the other hand, our conservative choice for the number density at which to compute the pQCD likelihood, in turn leads to a highly prior dominated pQCD posterior. The same phenomenon was seen in our model agnostic prior, where although the pQCD likelihood preferred lower pressures in the high-density regime, the effect was minimal compared to the astrophysical data. For a more complete exploration of the effect of this conservative pQCD input on our priors, see Appendix~\ref{App:pQCD}.

\begin{table}
\centering
\caption{Bayes Factor (BF) comparisons for each data stream, where A stands for anecdotal ($\text{BF} < 3$), M for moderate ($3 \leq \text{BF} < 10$), S for strong ($10 \leq \text{BF} < 30$), v.S for very strong ($30 \leq \text{BF} < 100$), and E for extreme ($\text{BF} \geq 100$).}\label{Tab:Evidences}
\begin{ruledtabular}
\begin{tabular}{lcc}
\textbf{Astrophysical} & Bayes Factor & Interpretation\\
\hline
Hadronic vs. Agnostic & $26.69$ &S \\
Hyperonic vs. Agnostic & $0.001$ & E, favors Agnostic \\
Quark vs. Agnostic & $0.017$ & v.S, favors Agnostic \\
\hline
\textbf{pQCD} & \\
\hline
Hadronic vs. Agnostic & $7.759$ &M \\
Hyperonic vs. Agnostic & $2.357$ &A\\
Quark vs. Agnostic & $1.343$ &A\\
\hline
\textbf{Combined} & \\
\hline
Hadronic vs. Agnostic & $207.1$ &E \\
Hyperonic vs. Agnostic & $0.003$ &E, favors Agnostic \\
Quark vs. Agnostic & $0.022$ &v.S, favors Agnostic \\

\end{tabular}
\end{ruledtabular}
\end{table}

The astrophysical data, one BNS Compact Binary Coalescence (CBC) gravitational wave event, two NICER mass and radius measurements and 105 pulsar mass observations, exhibit two notable effects on our separate prior analysis. Firstly, because the hyperonic models predict, on average, a larger tidal deformability, $\Lambda_\text{Hyperonic}(1.4\,M_\odot)= 819^{+300}_{-300}$, versus hadronic/quark models, $\Lambda_\text{Hadronic}(1.4\,M_\odot)= 513^{+300}_{-300}$, $\Lambda_\text{Quark}(1.4\,M_\odot)= 474^{+100}_{-100}$ (see Fig.~\ref{fig:corelations_informed}), the GW170817 data especially serves to constrain the possible hyperonic models, while having a minimal effect on the other two GMMs priors. On the other hand, the observations of massive NS, especially those above $2\,M_\odot$, largely restricts the quark and hyperonic models. A summary of these effects can be found in Table~\ref{Tab:Evidences}, which shows the Bayes factor of the 3 separate GMM priors relative to the model agnostic analysis. Astrophysical data alone is largely in favor of the hadronic prior, which by construction supports very high NS masses and moderate tidal deformabilities. As for the pQCD data alone, it has an anecdotal capacity only to distinguish between the three types of GMMs in our informed prior.

Regarding recent work with similar datasets~\cite{Biswas:2024hja,Fan:2023spm,Golomb:2024lds}, inferring the NS EoS jointly with NS population, our results are broadly consistent. For example, ref.~\cite{Biswas:2024hja} utilized a parametric EoS construction method (nuclear plus  piecewise polytropes), as well as chiral effective field theory ($\chi$EFT) results \cite{Meissner:2014lgi,Machleidt:2011zz,Hammer:2012id,Hebeler:2020ocj}, a recent mass-radius NICER measurement of J0437-4715 \cite{Choudhury:2024xbk} and neutron skin thickness measurements \cite{PREX:2021umo,CREX:2022kgg}, in addition to the astrophysical dataset and pQCD input we have considered. They similarly find that the pQCD input plays a minimal role in their inference, although it is important to note that they utilize the non-prior marginalized likelihood of ref.~\cite{Gorda:2022jvk}, and a lower integration condition based on $n_\text{TOV}$, the central density of a maximally massive NS, of each parametrized EoS, whereas we took a mean of our prior, see Appendix \ref{App:pQCD}. In terms of inferred NS properties, our results utilizing the model agnostic prior for $M_\text{max}$ and $\Lambda(1.4\,M_\odot)$ are in agreement with their findings. 
Ref.~\cite{Fan:2023spm} constructs three different (non)parametric EoS priors, including a GP-based one, and $\chi$EFT input, to constrain the maximum supported NS mass to $M_\text{max} = 2.25^{+0.08}_{-0.07}\,M_\odot$ (at 68.3\% credibility). Both refs.~\cite{Fan:2023spm,Biswas:2024hja} utilize the non-prior marginalized pQCD likelihood as well as ($\chi$EFT) results, and similar astrophysical datasets to ours. The tighter macroscopic NS observable constraints seen in refs.~\cite{Biswas:2024hja,Fan:2023spm} and this work, in relation to ref.~\cite{Golomb:2024lds} can somewhat be attributed to the level of flexibility of the EoS prior. As discussed in Section \ref{Sec:eos}, our ``model agnostic'' prior is closer to the ``model informed'' GMM construction of refs.~\cite{Landry:2019, Essick:2019ldf}. Therefore, the more flexible GP construction seen in ref.~\cite{Golomb:2024lds}, coupled with the lack of pQCD input, and the separate modeling of Galactic NS versus merging BNS populations, partly explain the larger confidence intervals, $M_\text{max} = 2.28^{+0.41}_{-0.21}\,M_\odot$ (at 90\% credibility). This further stresses the importance of complete unbiased NS population statistics. Current available NS observations, both with GW and EM, could also potentially suffer from selection effects, such as with GW detectors being more sensitive to higher mass events.

\section{Discussion and Conclusions}
\label{sec:conclu}

The behavior of extremely  cold dense nuclear matter remains an open question in nuclear astrophysics, not only due to modeling difficulties intrinsic to  such extreme environments as the cores of neutron stars (NSs), but also to the non-trivial connection between the macroscopic observables and underlying microphysics. 
We provide direct constraints on the neutron star equation of state (EoS), through the use of a novel analysis methodology, alongside a large set of astrophysical observations as well as  theoretical constraints. We construct two distinct priors by training Gaussian processes on  75 EoSs of cold nuclear matter. The model agnostic prior applies an equal-weighted treatment across the training set, while the model informed prior differentiates between hadronic, hyperonic, and quark EoS models. By treating the draws of each Gaussian mixture model prior, as a categorical distribution, our framework can be integrated in raw data parameter inference schemes. This is a scalable and efficient way of directly constraining the NS EoS in virtue of transparent non-parametric prior construction and ease of integration with existing Markov Chain Monte Carlo or nested sampling based parameter estimation pipelines.

Our astrophysical dataset, comprising the gravitational waveform of the GW170817 event, two NICER mass-radius posteriors and 105 NS mass observations, provides a two-front constraint on the EoS. In the low density regime, the tidal deformability influencing the gravitational waveform can distinguish  between our hyperonic and quark/hadronic priors. Conversely, in the high density regime, the larger tolerated masses of hadronic models are preferred by the common likelihood constructions for mass-radius or mass alone observations. As a result, higher pressures are preferred in the $\sim6 - 10 \rho_{\text{sat}}$ regime.

The inclusion of perturbative Quantum Chromodynamics (pQCD) calculations can potentially provide another accurate and independent avenue of narrowing down the NS EoS parameter space. The pQCD data alone predicts a lower range of pressures in the $\sim 6 - 10 \rho_{\text{sat}}$ regime, directly in contrast to the astrophysical data. While we find a very strong preference for the hadronic as opposed to quark prior using both datasets, it is important to note that the effect of the pQCD input is small in comparison to the astrophysical one. Furthermore, 
our model-informed prior, with GMMs trained separately on different EoS families, does result in a prior-dominated posterior, by construction. This effect is best seen in the different degrees of narrowing of the hadronic posterior, relative to its corresponding prior, which shows a larger information gain than in the case of the hyperonic/quark posteriors. However, this prior-dependence is used to further showcase the already observed, in the model agnostic prior, tension between the predicted pressures at high density between astrophysical likelihoods of Eqs.~\ref{eq:LikeNICER} and \ref{eq:LikeMass} versus the pQCD likelihood. Overall, the narrower priors coupled with the lower impact of the pQCD input, relative to massive pulsar observations, result in the observed Bayes factor difference, and subsequent hadronic preference. 

To conclude, our model agnostic prior, after the inclusion of astrophysical observations alone, within our analysis framework, provides an assumption-light and prior-transparent direct constraint on the NS equation of state. The advancements in pQCD calculations and the utilization of Gaussian process based extrapolations to the $\sim40\,n_{\text{sat}}$ density regime, can potentially offer another insight into the EoS of cold dense nuclear matter. Yet, the likelihood construction has to be carefully considered in the absence of complete NS population statistics and the model dependence of the extrapolation framework from NS interior densities to the pQCD regime.

\begin{acknowledgments}
We thank Leslie Wade,
Colm Talbot, and the anonymous referees for helpful comments and discussions.
IC acknowledges support through an EMJM scholarship funded by the European Union in the framework of the Erasmus+, Erasmus Mundus Joint Master in Astrophysics and Space Science – MASS. 
SR  was partially supported by the UK STFC grant ST/T000759/1 and by the Fermi National Accelerator Laboratory (Fermilab), a U.S.
Department of Energy, Office of Science, HEP User Facility. 
This work was performed in part at Aspen Center for Physics, which is supported by National Science Foundation grant PHY-2210452. 
SR  acknowledges CERN TH Department for its hospitality while this research was being carried out. 
This research was undertaken using the Computational Research, Engineering and Technology Environment (CREATE)~\cite{CREATEHPC} at King's College London, and Wilson Cluster at Fermilab. 
\end{acknowledgments}

\appendix

\section{Equation of state training catalog}
\label{sec:EOSsample}
We select a large representative sample of 75 equations of state of cold neutron stars from the CompOSE database~\footnote{ \url{https://compose.obspm.fr}} \cite{Typel:2013rza,Oertel:2016bki,CompOSECoreTeam:2022ddl}. Table~\ref{tab:EOSsample} lists all the models used for training, alongside detailed information on particle content as well as calculation methods. 
Calculation methods for homogeneous matter in the core are shown in the third column, which include: Microscopic calculations - EoS based on ab initio calculations of nuclear matter; Non-relativistic density functional - baryonic interaction given by Skyrme-like  or Gogny-type interactions; Relativistic density functional - covariant density functional theory, models generally in the mean-field approximation. Similarly, the treatment of the inhomogeneous matter in the crust is given in the fourth column: Crust-core matched - non-unified models; Unified EoS - the same nuclear interactions describe both crust and core; Not treated - EoS lacks direct crust description; SNA models - use the single nucleus approximation for inhomogeneous matter at nonzero temperature; NSE models - use statistical equilibrium approach to describe inhomogeneous matter at nonzero temperature.
For more information, refer to the CompOSE manual~\footnote{\url{https://compose.obspm.fr/manual/}}.

\begin{longtable*}{|c|c|c|c|c|}
    \caption{Catalog of Equations of State} \label{tab:EOSsample} \\
    \hline
    \multicolumn{5}{|c|}{\textbf{Nucleonic Models}} \\ 
    \hline
    \endfirsthead
    
    \hline
    \textbf{Name} & \textbf{Particles} & \textbf{Homogeneous} & \textbf{Inhomogeneous} & \textbf{Ref.} \\  \hline
    \endhead
    
    \hline \multicolumn{5}{r}{\textit{Continued on next page}} \\ \hline
    \endfoot
    
    \hline
    \endlastfoot
    
     \hline
    \textbf{Name} & \textbf{Particle} & \textbf{Calculation Method} & \textbf{Inhomogeneous } & \textbf{Reference} \\ 
    \textbf{} & \textbf{Content} & \textbf{Homogeneous Matter} & \textbf{Matter} & \textbf{} \\
    \hline   
    APR(APR) & npe$\mu$ & Microscopic calculations & Crust-core matched & \cite{Akmal:1998cf,Douchin:2001sv,Baym:1971pw} \\ 
    BBB(BHF-BBB2) & npe & Microscopic calculations & Crust-core matched & \cite{Douchin:2001sv,Baym:1971pw,Baldo:1997ag} \\ 
    BL(chiral) & npe$\mu$N & Microscopic calculations & Unified EoS & 
   \cite{Bombaci:2018ksa,Carreau:2019zdy}  \\
    BSK20 & \multirow{2}{*}{npe$\mu$} & \multirow{2}{*}{Non-relativ. density functional} & \multirow{2}{*}{Unified EoS} & \multirow{2}{*}{\cite{Goriely:2010bm}}\\ 
    BSK21 &  &  &  & \\ 
    PCP(BSK22) & npe$\mu$N & Non-relativ. density functional & Unified EoS & \multirow{4}{*}{\cite{Audi:2017asy,Pearson:2018tkr,Pearson:2020bxz,Pearson:2022vep,Allard:2021rrt,Goriely:2013xba,Perot:2019gwl,Welker:2017eja,Xu:2012uw}} \\ 
    PCP(BSK24) & npe$\mu$N & Non-relativ. density functional & Unified EoS & \\ 
    PCP(BSK25) & npe$\mu$N & Non-relativ. density functional & Unified EoS & \\ 
    PCP(BSK26) & npe$\mu$N & Non-relativ. density functional & Unified EoS& \\ 
    ENG & npe & Non-relativ. density functional & Crust-core matched & \cite{Engvik:1995gn}\\
    KDE0v & npe$\mu$ & Non-relativ. density functional & Unified EoS & \multirow{2}{*}{\cite{Danielewicz:2008cm,Gulminelli:2015csa,Agrawal:2005ix}}\\ 
    KDE0v1 & npe$\mu$ & Non-relativ. density functional & Unified EoS & \\ 
    MPA1 & npe & Relativistic density functional & Not treated & \cite{Muther:1987xaa}\\
    RS & npe$\mu$ & Non-relativ. density functional & Unified EoS & \cite{Friedrich:1986zza}\\
    SK255 & \multirow{2}{*}{npe} & \multirow{2}{*}{Non-relativ. density functional} & \multirow{2}{*}{Not treated} & \multirow{2}{*}{\cite{Agrawal:2003xb}}\\
    SK272 &  &  &  & \\
    SKI2 & \multirow{4}{*}{npe} & \multirow{4}{*}{Non-relativ. density functional} & \multirow{4}{*}{Not treated} & \multirow{4}{*}{\cite{Reinhard:1995zz}}\\
    SKI3 &  &  &  & \\
    SKI4 &  &  &  & \\
    SKI5 &  &  &  & \\
    SKI6 & npe & Non-relativ. density functional & Not treated & \cite{nazarewicz1996structure} \\
    SKMP & npe &  &  & \cite{Bennour:1989zz}\\
    SKOP & npe & Non-relativ. density functional & Not treated & \cite{Reinhard:1999ut}\\
    SLY & npe$\mu$ & Non-relativ. density functional & Unified  EoS & \cite{Douchin:2001sv}\\
    SLY2 & \multirow{2}{*}{npe$\mu$} & \multirow{2}{*}{Non-relativ. density functional} & \multirow{2}{*}{Unified  EoS} & \multirow{2}{*}{\cite{Chabanat:1995tea}}\\
    SLY9 & &  &  & \\
    SLY230a & npe$\mu$ & Non-relativ. density functional & Unified  EoS & \cite{Chabanat:1997qh}\\
    SPG(M1) & npe$\mu$N & Relativistic density functional & Unified  EoS & \multirow{5}{*}{\cite{Pearson:2018tkr,Gogelein:2007qa,Scurto:2024ekq}}\\
    SPG(M2) & npe$\mu$N & Relativistic density functional & Unified  EoS& \\
    SPG(M3) & npe$\mu$N & Relativistic density functional & Unified  EoS & \\
    SPG(M4) & npe$\mu$N & Relativistic density functional & Unified  EoS & \\
    SPG(M5) & npe$\mu$N & Relativistic density functional & Unified  EoS & \\
    \hline
    
    \multicolumn{5}{|c|}{\textbf{Models with hyperons (\& Delta-Resonances)}} \\\hline
    \textbf{Name} & \textbf{Particle} & \textbf{Calculation Method} & \textbf{Inhomogeneous} & \textbf{Reference} \\ 
    \textbf{} & \textbf{Content} & \textbf{Homogeneous Matter} & \textbf{Matter} & \textbf{} \\
    \hline
    DNS(CMF) & npe$\mu$Bs & Relativistic density functional & Crust-core matched & \cite{Dexheimer:2008ax,Dexheimer:2015qha,Schurhoff:2010ph,Dexheimer:2017nse,Danielewicz:2008cm,Friedrich:1986zza,Gulminelli:2015csa} \\ 
    DS(CMF)-1 & npeNBs & Relativistic density functional & Crust-core matched & \\ 
    DS(CMF)-2 & npeN & Relativistic density functional & Crust-core matched &  \\ 
    DS(CMF)-3 & npeNBs & Relativistic density functional & Crust-core matched & \\ 
    DS(CMF)-4 & npeN & Relativistic density functional & Crust-core matched &  \cite{Bennour:1989zz,Gulminelli:2015csa,Dexheimer:2008ax}\\ 
    DS(CMF)-5 & npeNBs & Relativistic density functional & Crust-core matched &  \cite{Dexheimer:2017nse,Dexheimer:2009hi,Dexheimer:2020rlp}\\ 
    DS(CMF)-6 & npeN & Relativistic density functional & Crust-core matched &  \\ 
    DS(CMF)-7 & npeNBsD & Relativistic density functional & Crust-core matched &  \\ 
    DS(CMF)-8 & npeND & Relativistic density functional & Crust-core matched &  \\ 
    H4 & npeBs & Relativistic density functional & Crust-core matched & \cite{Lackey:2005tk}\\
    OPGR(DDHdeltaY4) & npeBs & Relativistic density functional & Crust-core matched & \cite{Gaitanos:2003zg,Oertel:2014qza,Grill:2014aea,Douchin:2001sv}\\ 
    OPGR(GM1Y4) & npeBs & Relativistic density functional & Crust-core matched & \multirow{3}{*}{\cite{Glendenning:1991es,Oertel:2014qza,Douchin:2001sv}}\\ 
    OPGR(GM1Y5) & npeBs & Relativistic density functional & Crust-core matched & \\ 
   OPGR(GM1Y6) & npeBs & Relativistic density functional & Crust-core matched & \\ 
    R(DD2YDelta) 1.1-1.1 & npeNBsDelt & Relativistic density functional & Crust-core matched & \multirow{3}{*}{\cite{Typel:2009sy,Raduta:2020fdn,Raduta:2022elz,Vinas:2021vmv}}\\ 
    R(DD2YDelta) 1.2-1.1 & npeNBsDelt & Relativistic density functional & Crust-core matched & \\ 
    R(DD2YDelta) 1.2-1.3 & npeNBsDelt & Relativistic density functional & Crust-core matched & \\ 
    \hline
    
    \multicolumn{5}{|c|}{\textbf{Hybrid (Quark-Hadron) Models}} \\\hline
    \textbf{Name} & \textbf{Particle} & \textbf{Calculation Method} & \textbf{Inhomogeneous} & \textbf{Reference} \\ 
    \textbf{} & \textbf{Content} & \textbf{Homogeneous Matter} & \textbf{Matter} & \textbf{} \\
    \hline
    ALF2 & \multirow{2}{*}{npq} & \multirow{2}{*}{Microscopic calculations} & \multirow{2}{*}{Not treated} & \multirow{2}{*}{\cite{alford2005hybrid}} \\ 
    ALF4 &  &  &  &  \\
    BHK(QHC18) & npeNq & Microscopic calculations & SNA models & \cite{Akmal:1998cf,Togashi:2017mjp,baym2018hadrons} \\
    BFH(QHC19-A) & npeNq & Microscopic calculations & SNA models & \multirow{4}{*}{\cite{Togashi:2017mjp,baym2018hadrons,Baym:2019iky}} \\ 
    BFH(QHC19-B) & npeNq & Microscopic calculations & SNA models &  \\ 
    BFH(QHC19-C) & npeNq & Microscopic calculations & SNA models &  \\ 
    BFH(QHC19-D) & npeNq & Microscopic calculations & SNA models &  \\ 

    DS(CMF)-1 Hybrid & npeNBsq & Relativistic density functional & Crust-core matched & \\ 
    DS(CMF)-2 Hybrid & npeNq & Relativistic density functional & Crust-core matched &\\ 
    DS(CMF)-3 Hybrid & npeBsq & Relativistic density functional & Not treated & \cite{Danielewicz:2008cm,Gulminelli:2015csa,Dexheimer:2008ax}\\ 
    DS(CMF)-4 Hybrid & npeq & Relativistic density functional & Not treated & \cite{Dexheimer:2017nse,Friedrich:1986zza,Bennour:1989zz,Dexheimer:2020rlp,Dexheimer:2009hi} \\ 
    DS(CMF)-5 Hybrid & npeBsq & Relativistic density functional & Not treated & \cite{Clevinger:2022xzl,Dexheimer:2018dhb}\\ 
    DS(CMF)-6 Hybrid & npeq & Relativistic density functional & Not treated & \\ 
    DS(CMF)-7 Hybrid & npeBsDeltq & Relativistic density functional & Not treated & \\ 
    DS(CMF)-8 Hybrid & npeDeltq & Relativistic density functional & Crust-core matched & \\ 
    
    JJ(VQCD(APR)), soft & npeNq & Holographic models & Crust-core matched & \multirow{3}{*}{\cite{Akmal:1998cf,Jokela:2018ers,Ishii:2019gta,Ecker:2019xrw,Jokela:2020piw}}\\ 
    JJ(VQCD(APR)), intermediate& npeNq & Holographic models & Crust-core matched & \\ 
    JJ(VQCD(APR)), stiff& npNq & Holographic models & Crust-core matched & \\ 

    KBH(QHC21-A) & npeNq & Microscopic calculations & SNA models & \multirow{4}{*}{\cite{Togashi:2017mjp,Kojo:2021wax,Drischler:2020fvz}}\\ 
    KBH(QHC21-B) & npeNq & Microscopic calculations & SNA models & \\
    KBH(QHC21-C) & npeNq & Microscopic calculations & SNA models & \\
    KBH(QHC21-D) & npeNq & Microscopic calculations & SNA models & \\
    
    OOS(DD2-FRG), 2 flavors & npeNq & Relativistic density functional & NSE models & \multirow{6}{*}{\cite{Hempel:2009mc,Typel:2009sy,Otto:2019zjy,Otto:2020hoz}}\\ 

    OOS(DD2-FRG), 2+1 flavors & npeNqqs & Relativistic density functional & NSE models & \\ 
    
    OOS(DD2-FRG), w/& \multirow{2}{*}{npeNq} & \multirow{2}{*}{Relativistic density functional} & \multirow{2}{*}{NSE models} & \\ 
    vect. interactions (2 flavors) &  &  &  & \\

    OOS(DD2-FRG), w/& \multirow{2}{*}{npeNqs} & \multirow{2}{*}{Relativistic density functional} & \multirow{2}{*}{NSE models} & \\ 
    vect. interactions (2+1 flavors) &  &  &  & \\
    
    \hline

\end{longtable*}

In terms of the hyper-prior distributions used for the cross validation likelihood of Eq.~\ref{eq:crossvalid}, the RBF length scale prior is a linear distribution, $[0.1,5]$, the model variance prior is a log-linear distribution,  $[0.1,1]$, and the noise variance prior is linear in $[0.02,1]$. As for the priors of the individual EoS modeling GPs, the same hyper-prior is utilized. We vary the \emph{direct tuning parameter}, or temperature $T$, between the two hyper-priors. This parameter was introduced in ref.~\cite{Landry:2020vaw} as a measure of model agnosticism, which effectively acts as a penalty term when joining GPs to construct the overarching process. Since we aim to construct faithful GP representations of tabulated EoS data, we take $T=1$, but when joining multiple GPs, we penalize every GP's contribution slightly, $T=10$, so that we allow for more model variance in our model agnostic prior. For the model informed prior, we again do not penalize any GP, so $T=1$, and utilize the same RBF kernel hyper-prior as before.

\begin{figure}
    \centering
    \includegraphics[width=\linewidth]{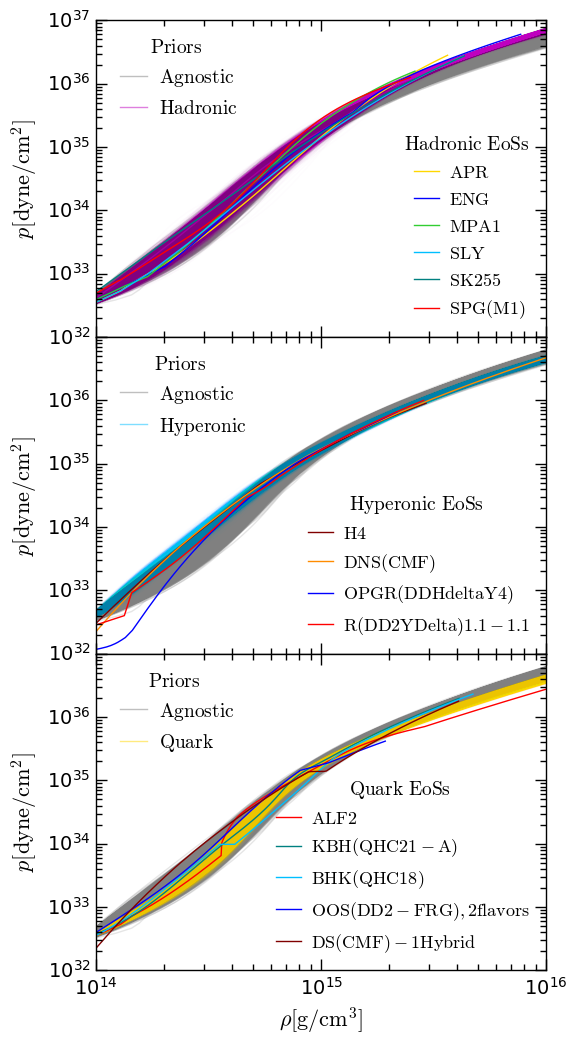}
    \caption{Comparison of agnostic (gray region) and informed priors (magenta for hadronic,  light blue for hyperonic and yellow for quark), with respect to some of the tabulated EoSs used for training. For reference, the mean $\rho_\text{max}=1.70\times10^{15}\g/\cm^3$ for the model agnostic prior, $\rho_\text{max}=1.69\times10^{15}\g/\cm^3$ for hadronic prior, $\rho_\text{max}=1.59\times10^{15}\g/\cm^3$ for hyperonic and $\rho_\text{max}=1.64\times10^{15}\g/\cm^3$ for quark, where $\rho_\text{max}$ is the central density of a maximally massive NS.}
    \label{fig:reprEoS}
\end{figure}

For a visual reference on how the tabulated EoSs used for training inform the resulting priors, see Figure~\ref{fig:reprEoS}. The model agnostic and model informed priors are plotted in pressure versus  density as well as some representative tabulated EoSs used for training. It is important to mention, that by $\sim 1 \times 10^{15} \g/\cm^3$, with the exact value depending on the EoS, we reach the energy density corresponding to the maximum NS mass, $M_{\text{max}}$, i.e. the astrophysically relevant range. Above this range, the priors become more disjoint, with the most relevant difference being between some of the hadronic EoSs used for training and the agnostic prior. This effect is more pronounced in the derivative of the EoS, this is, in both $c_s(\rho)$, Fig.~\ref{fig:draws}, and $\phi(p)$, Fig.~\ref{fig:GMM}, representations. This is mainly due to the fact  that the variance of each EoS's GP grows dramatically at high pressures or densities which translates into a high variance in the GMM, the reason for this being the resampling scheme itself. Specifically, some tabulated EoSs end in a pressure before the last point considered for the GP/GMM construction, so the GP effectively extrapolates the EoS, hence the estimated errors grow. Even though this happens only to some EoSs, it propagates to the resulting GMM. Considering the amount of EoSs used for training, and their varied phenomenology, it is to be expected that some of them would lie outside the 10000 draws of the prior agnostic construction at very high densities. This again reflects the versatility of choosing how closely the GMM follows the underlying GPs, based on the direct tuning parameter, but also the importance of carefully dealing with any extrapolations of the training data. However, as mentioned above, this effect is minimal in the astrophysically relevant range, with the resulting priors being good representation of the training data.

\section{NS mass observation catalog} 
\label{App:NSobs}

The neutron star mass observations utilized in our analysis are listed in Table~\ref{tab:NScatalog}. These 105 NS observations are all events up to April 2023 that offer direct mass measurements. Naturally, the NICER events PSR J0030+045 and PSR J0740+6620, as well as LIGO's GW170817 BNS CBS are excluded from this table, since we treat them separately.

\begin{longtable}{|l|c|c|c|}
    \caption{Catalog of 105 NS with direct mass measurements. WD stands for white dwarf, BW for black widow millisecond pulsar system, RB for Redback millisecond pulsar, INS for isolated neutron star; MS for main sequence star, HMXB, high mass X-ray binary; LMXB, low mass X-ray binary. This was adapted from ref.~\cite{Fan:2023spm}.} \label{tab:NScatalog} \\ 
    \hline
    \textbf{ID} & \textbf{Type} & \textbf{Mass}  [$M_\odot$]  & \textbf{Ref.} \\ 
    \hline
    \endfirsthead

    \hline
    \textbf{ID} & \textbf{Type} & \textbf{Mass} [$M_\odot$] & \textbf{Ref.} \\ 
    \hline
    \endhead

    \hline
    \multicolumn{4}{|r|}{\textit{Continued on next column}} \\
    \hline
    \endfoot

    \hline
    \endlastfoot

    J0453+1559  & NS-NS & 1.559$\pm$0.005 & \cite{Martinez:2015mya} \\
    J0453+1559 comp.  & NS-NS & 1.174$\pm$0.004 & \cite{Martinez:2015mya} \\
    J0509+3801  & NS-NS & 1.34$\pm$0.08 & \cite{Lynch:2018zxo} \\
    J0509+3801 comp.  & NS-NS &  1.46$\pm$0.08 & \cite{Lynch:2018zxo} \\
    J0514-4002A  & NS-NS & $1.25^{+0.05}_{-0.06}$ & \cite{Ridolfi:2019wgs} \\
    J0514-4002A comp.  & NS-NS & $1.22^{+0.06}_{-0.05}$ & \cite{Ridolfi:2019wgs} \\
    J0737-3039A  & NS-NS & $1.338185^{+0.000012}_{-0.000014}$ & \cite{Swiggum:2022xlb} \\
    J0737-3039B  & NS-NS & $1.248868^{+0.000013}_{-0.000011}$ & \cite{Swiggum:2022xlb} \\
    B1534+12  & NS-NS & 1.3330$\pm$0.0002 & \cite{Fonseca:2014qla} \\
    B1534+12 comp.  & NS-NS &  1.3455$\pm$0.0002 & \cite{Fonseca:2014qla} \\
    J1756-2251  & NS-NS & 1.341$\pm$0.007 & \cite{Ferdman:2014rna} \\
    J1756-2251 comp.  & NS-NS & 1.230$\pm$0.007 & \cite{Ferdman:2014rna} \\
    J1757-1854  & NS-NS & 1.3406$\pm$0.0005 & \cite{Cameron:2022soa} \\
    J1757-1854 comp.  & NS-NS &  1.3922$\pm$0.0005 & \cite{Cameron:2022soa} \\
    J1807-2500B  & NS-NS &  1.3655$\pm$0.0021 & \cite{Lynch:2011aa} \\
    J1807-2500B comp.  & NS-NS & 1.2064$\pm$0.0020 & \cite{Lynch:2011aa} \\
    J1829+2456  & NS-NS & 1.306$\pm$0.007 & \cite{Haniewicz:2020jro} \\
    J1829+2456 comp.  & NS-NS &  1.299$\pm$0.007 & \cite{Haniewicz:2020jro} \\
    J1906+0746  & NS-NS & 1.291$\pm$0.011 & \cite{vanLeeuwen:2014sca} \\
    J1906+0746 comp.  & NS-NS &  1.322$\pm$0.011 & \cite{vanLeeuwen:2014sca} \\
    J1913+1102  & NS-NS & 1.62$\pm$0.03 &\cite{Ferdman:2020huz} \\
    J1913+1102 comp.  & NS-NS & 1.27$\pm$0.03 & \cite{Ferdman:2020huz} \\
    B1913+16  & NS-NS &  1.4398$\pm$0.0002 & \cite{Weisberg:2010zz} \\
    B1913+16 comp.  & NS-NS &  1.3886$\pm$0.0002 & \cite{Weisberg:2010zz} \\
    B2127+11C & NS-NS &  1.358$\pm$0.010 & \cite{Jacoby:2006dy} \\
    B2127+11C comp.  & NS-NS & 1.354$\pm$0.010 & \cite{Jacoby:2006dy} \\
    J0337+1715  & NS-WD & 1.4359$\pm$0.0003 & \cite{Archibald:2018oxs} \\
    J0348+0432  & NS-WD & 2.01$\pm$0.04 & \cite{Antoniadis:2013pzd} \\
    J0437-4715  & NS-WD &  1.44$\pm$0.07 & \cite{Reardon:2015kba} \\
    J0621+1002  & NS-WD & $1.53^{+0.10}_{-0.20}$ & \cite{kasian2012radio} \\
    J0751+1807  & NS-WD & 1.64$\pm$0.15 & \cite{EPTA:2016ndq} \\
    J0955-6150  & NS-WD &  1.71$\pm$0.02 & \cite{Serylak:2022kna} \\
    J1012+5307  & NS-WD & 1.72$\pm$0.16 & \cite{MataSanchez:2020pys} \\
    J1017-7156  & NS-WD & 2.0$\pm$0.8 & \cite{Reardon:2021gko} \\
    J1022+1001  & NS-WD & 1.44$\pm$0.44 & \cite{Reardon:2021gko} \\
    J1125-6014  & NS-WD & 1.5$\pm$0.2 & \cite{Reardon:2021gko}\\
    J1141-6545  & NS-WD & 1.27$\pm$0.01 &\cite{NANOGrav:2017wvv}\\
    B1516+02B  & NS-WD & 2.08$\pm$0.19 &\cite{Freire:2007jd}\\
    J1528-3146  & NS-WD & $1.61^{+0.14}_{-0.13}$ & \cite{Berthereau:2023aod} \\
    J1600-3053  & NS-WD & $2.3^{+0.7}_{-0.6}$ & \cite{NANOGrav:2017wvv} \\
    J1614-2230  & NS-WD & 1.908$\pm$0.016 & \cite{NANOGrav:2017wvv}\\
    J1713+0747  & NS-WD &  1.35$\pm$0.07 & \cite{NANOGrav:2017wvv} \\
    J1738+0333  & NS-WD & $1.47^{+0.7}_{-0.6}$ & \cite{Antoniadis:2012vy} \\
    J1741+1351  & NS-WD & $1.14^{+0.43}_{-0.25}$ & \cite{NANOGrav:2017wvv} \\
    J1748-2446am  & NS-WD & $1.649^{+0.037}_{-0.11}$ & \cite{Andersen:2018nsx} \\
    B1802-07  & NS-WD & $1.26^{+0.08}_{-0.17}$ & \cite{Thorsett:1998uc} \\
    J1802-2124  & NS-WD & 1.24$\pm$0.11 & \cite{Ferdman:2010rk} \\
    J1811-2405  & NS-WD & $2.0^{+0.8}_{-0.5}$ & \cite{Ng:2020uck} \\
    B1855+09  & NS-WD & $1.37^{+0.13}_{-0.10}$ & \cite{NANOGrav:2017wvv} \\
    J1909-3744  & NS-WD & 1.492$\pm$0.014 & \cite{Liu:2020hkx} \\
    J1911-5958A  & NS-WD & 1.34$\pm$0.08 & \cite{Bassa:2006mj} \\
    J1918-0642  & NS-WD & 1.29$\pm$0.1 & \cite{NANOGrav:2017wvv} \\
    J1946+3417  & NS-WD &  1.828$\pm$0.022 &\cite{Barr:2016vxv} \\
    J1949+3106  & NS-WD & $1.34^{+0.17}_{-0.15}$ & \cite{Zhu:2019oax}\\
    J1950+2414  & NS-WD & 1.496$\pm$0.023 & \cite{Zhu:2019oax} \\
    J2043+1711  & NS-WD & $1.38^{+0.12}_{-0.13}$ & \cite{NANOGrav:2017wvv}\\
    J2045+3633  & NS-WD & 1.251$\pm$0.021 & \cite{McKee:2020pzp}\\
    J2053+4650  & NS-WD & $1.40^{+0.21}_{-0.18}$ & \cite{Berezina:2017vts}\\
    J2222-0137  & NS-WD & 1.831$\pm$0.010 & \cite{Guo:2021bqa}\\
    J2234+0611  & NS-WD & $1.353^{+0.014}_{-0.017}$ & \cite{Stovall:2018rvy}\\
    B2303+46  & NS-WD & $1.30^{+0.13}_{-0.46}$ & \cite{Thorsett:1998uc} \\
    J0952-0607  & BW & 2.35$\pm$0.17 & \cite{Romani:2022jhd}\\
    J1301+0833  & BW & $1.60^{+0.22}_{-0.25}$ & \cite{Kandel:2022qor}\\
    J1311-3430  & BW & 2.22$\pm$0.1 &\cite{Kandel:2022qor} \\
    J1555-2908  & BW & $1.67^{+0.07}_{-0.05}$ & \cite{Kennedy:2022zml} \\
    J1653-0158  & BW & 2.15$\pm$0.16 &\cite{Kandel:2022qor}\\
    J1810+1744  & BW & 2.11$\pm$0.04 & \cite{Kandel:2022qor}\\
    J1959+2048  & BW & 1.81$\pm$0.07 & \cite{Clark:2023owb} \\
    3FGL J0212.1+5320  & RB & $1.85^{+0.32}_{-0.26}$ & \cite{Shahbaz:2017awv} \\
    3FGL J0427.9-6704  & RB & $1.86^{+0.11}_{-0.10}$ &\cite{Strader:2016qpu}\\
    2FGL J0846.0+2820  & RB &  1.96$\pm$0.41 & \cite{Swihart:2017yaz} \\
    J1023+0038  & RB & $1.65^{+0.19}_{-0.16}$ &\cite{Strader:2018qbi} \\
    1FGL J1417.7-4407  & RB & $1.62^{+0.43}_{-0.17}$ & \cite{Swihart:2018nur} \\
    J1723-2837  & RB & $1.22^{+0.26}_{-0.20}$ & \cite{Strader:2018qbi} \\
    4FGL J2039.5-5617  & RB & $1.3^{+0.155}_{-0.10}$ & \cite{Clark:2020hbv} \\
    3FGL J2039.6-5618  & RB & $2.04^{+0.37}_{-0.25}$ & \cite{Strader:2018qbi} \\
    J2129-0429  & RB & 1.74$\pm$0.18 & \cite{Bellm:2015dfa}\\
    J2215+5135  & RB & $2.28^{+0.10}_{-0.09}$ & \cite{Kandel:2020xha}\\
    J2339-0533  & RB & 1.47$\pm$0.09 &\cite{Kandel:2020xha} \\
    J0045-7319  &  NS-MS & 1.58$\pm$0.34 & \cite{bell1995psr} \\
    J1903+0327  &  NS-MS & $1.666^{+0.010}_{-0.012}$ & \cite{NANOGrav:2017wvv} \\
    4U1538-522  & HMXB &  1.02$\pm$0.17 & \cite{Falanga:2015mra}\\
    4U1700-377  & HMXB & 1.96$\pm$0.19 &\cite{Falanga:2015mra} \\
    Cen X-3  & HMXB &  1.57$\pm$0.16 &\cite{Falanga:2015mra} \\
    EXO 1722-363  & HMXB & 1.91$\pm$0.45 & \cite{Falanga:2015mra} \\
    Her X-1  & HMXB & 1.07$\pm$0.36 & \cite{Rawls:2011jw} \\
    J013236.7+303228  & HMXB &  2.0$\pm$0.4 & \cite{Bhalerao:2012xe} \\
    LMC X-4  & HMXB &  1.57$\pm$0.11 & \cite{Bellm:2015dfa} \\
    OAO 1657-415  & HMXB &  1.74$\pm$0.3 & \cite{Bellm:2015dfa} \\
    SAX J1802.7-2017  & HMXB &  1.57$\pm$0.25 & \cite{Bellm:2015dfa}\\
    SMC X-1  & HMXB &  1.21$\pm$0.12 & \cite{Bellm:2015dfa} \\
    Vela X-1  & HMXB & 2.12$\pm$0.16 & \cite{Bellm:2015dfa} \\
    XTE J1855-026  & HMXB &  1.41$\pm$0.24 & \cite{Bellm:2015dfa} \\
    2S 0921-630  &  LMXB &  1.44$\pm$0.1 &\cite{Steeghs:2007cm}\\
    4U 1608-52  &  LMXB & $1.57^{+0.30}_{-0.29}$ & \cite{Ozel:2015fia}\\
    4U1702-429  &  LMXB & 1.9$\pm$0.3 & \cite{Nattila:2017wtj}\\
    4U 1724-207  &  LMXB & $1.81^{+0.25}_{-0.37}$ & \cite{Ozel:2015fia} \\
    4U 1820-30  &  LMXB & $1.77^{+0.25}_{-0.28}$ & \cite{Ozel:2015fia} \\
    Cyg X-2  &  LMXB &  1.71$\pm$0.21 & \cite{Casares:2009vq} \\
    KS 1731-260  &  LMXB & $1.61^{+0.35}_{-0.37}$ & \cite{Ozel:2015fia} \\
    EXO 1745-248  &  LMXB & $1.65^{+0.21}_{-0.31}$ & \cite{Ozel:2015fia} \\
    SAX J1748.9-2021  &  LMXB & $1.81^{+0.25}_{-0.37}$ & \cite{Ozel:2015fia} \\
    X 1822-371  &  LMXB &  1.96$\pm$0.36 & \cite{Munoz-Darias:2005dwo}\\
    XTE J2123-058  &  LMXB &  1.53$\pm$0.42 &\cite{Gelino2002}\\
\hline
\end{longtable}

\section{GW170817 analysis} 
\label{App:gw170817}

\begin{figure*}
    \centering
    \includegraphics[width=\textwidth]{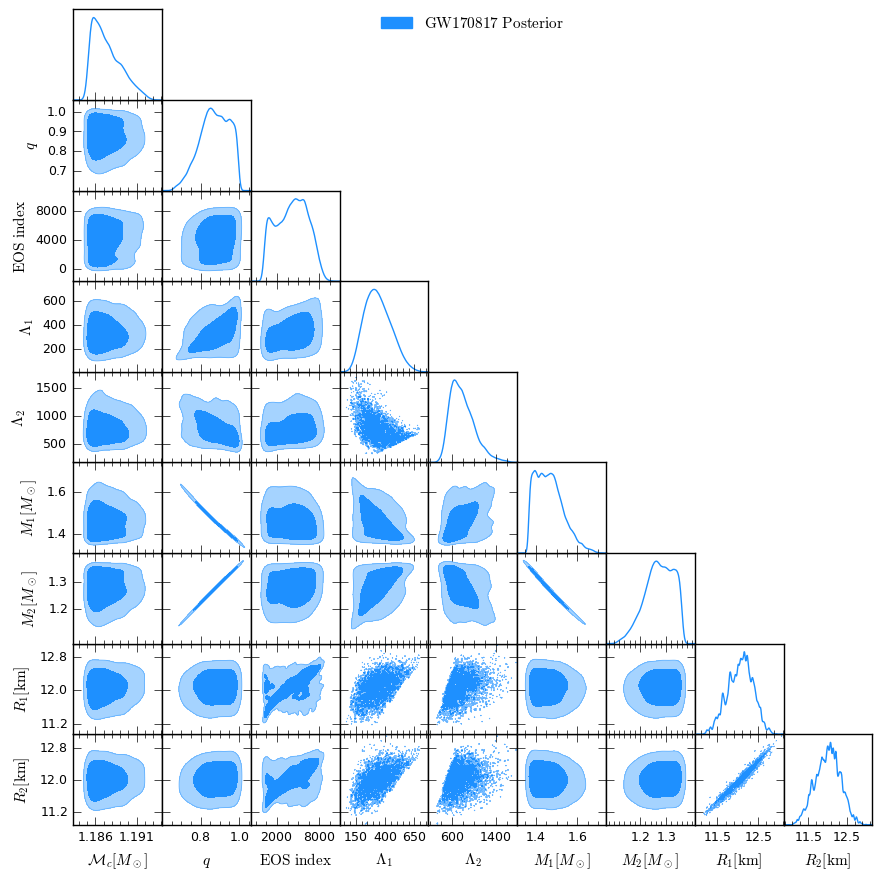}
    \caption{Corner plot of the source parameters of the BNS CBC GW170817 event resulting from our modified analysis scheme. Besides the spatio-temporal localization, along with the six spins, only the chirp mass, $\mathcal{M}_c = 1.18741^{+0.00090}_{-0.0023}[M_\odot]$, mass ratio, $q = 0.873^{+0.10}_{-0.056}$ and the EoS Index (stand-in for which EoS from the prior is assumed at a given likelihood computation, see Fig.~\ref{fig:agn_result_gw}) were directly sampled. The component masses, $M_1 = 1.465^{+0.036}_{-0.091}[M_\odot]$ and $M_2 = 1.274^{+0.074}_{-0.038}[M_\odot]$, tidal deformabilities, $\Lambda_1 = 339^{+90}_{-100}$ and $\Lambda_2 = 791^{+100}_{-300}$, along with the component radii , $R_1 = 12.060^{+0.340}_{-0.280}[\km]$ and $R_2 = 12.010^{+0.360}_{-0.290}[\km]$ are all derived parameters.}
    \label{fig:gw17cp_post}
\end{figure*}

Our analysis assumes that the GW170817 signal comes from a BNS CBC, located at the sky position of the GRB 170817A event, with the time of coalescence estimated at $1187008882.43$ geocentric time~\cite{LIGOScientific:2017zic}. We also assume that both neutron stars are governed by the same physical equation of state,  
and the individual spins are typical of BNS systems. A summary of the priors can be found in Table~\ref{tab:Priors}.

The sky-position of the event is set to a very narrow Gaussian prior, found in ref.~\cite{LIGOScientific:2017ync}, given by the results of the $1.7\s$ delayed EM counterpart to GW170817, measured by the Fermi Gamma-ray Burst Monitor \cite{Meegan:2009} and designated GRB 170817A. Lastly, the chirp mass of the system has been constrained to the measured value in the detection analysis \cite{LIGOScientific:2017vwq}. This was implemented through setting a ``uniform in components'' chirp mass prior of $\sim 1.1975\,M_\odot$. This means that the prior distribution of the chirp mass and mass ratios, which are the parameters being varied, are constructed using uniform distributions on the components masses, but under the constraint that they result in localized distribution on the chirp mass and mass ratio. This allows us to specifically impose a very tight prior on the chirp mass, in line with previous EoS analyses \cite{LIGOScientific:2018hze}, without artificially imposing tight constraints on the component masses. We also note that we allow for the mass ratio prior range to be quite wide, see Table~\ref{tab:Priors}.

Regarding the spins, it was found in ref.~\cite{LIGOScientific:2018hze} that the choice of prior had a strong impact on the spin inferences, and in turn on the mass ratio (mass ratio - effective spin degeneracy). Because of this, together with the fact that some waveform approximants impose aligned spins, we found no reason to further restrict the spin prior, through assumptions on the existence of precession effects. Thus, we restrict the dimensionless spin magnitudes $a_1, a_2 < 0.05$, consistent with the population of BNS systems (see e.g., ref.~\cite{Tauris:2017omb}). 
In the observer frame, the other four spin parameters are: the two tilt angles between the components' spins and  orbital angular momentum $\theta_1, \theta_2$ and finally the $2$ spin vectors which describe the azimuthal angle separation $\delta\phi$ and the cone of precession around the system's angular momentum $\phi_{JL}$. Further information about this spin basis selection, as opposed to the standard source frame, $S_{x, y, z}$ can be found in ref.~\cite{Ashton:2018jfp}.

Lastly, the exact temporal localization of the signal, specifically the parameter governing the time of coalescence, is of particular importance. It is a leading order term whose constraints directly impact all other inferred parameters. Also noteworthy, is that choosing a wide prior on the time of coalescence (say a uniform prior  ${\cal O}(10\s)$, as in ref.~\cite{LIGOScientific:2017vwq}) leads to two distinct peaks in the posterior of the time of coalescence. Considering that all waveform approximants used to model BNS systems are cut off at the moment of merger, this directly influences how much of the very high-frequency, late inspiral information, is extracted. Although a prior that is much broader than the relevant timeframes (milliseconds in the case of BNSs) is perhaps ideal for a detection analysis, this might not be the case if one is to utilize the already confirmed detection as a BNS source to infer EoS information. As a result, we use a more informative, and thus narrower, Gaussian prior on the estimated time of coalescence.

As mentioned in the main text, to speed up sampling, we utilize the relative binning likelihood \cite{Zackay:2018qdy}. Briefly, instead of computing the model at every likelihood evaluation throughout the frequency space, the relative binning method computes the ratio between the model and a fiducial waveform at specified bin edges. This is a faster method, since this ratio varies almost linearly with small parameter changes, and the bin selection mediates the accuracy. For a more detailed explanation of the implementation of this likelihood in \texttt{Bilby}, see ref.~\cite{Krishna:2023bug}.

Figure~\ref{fig:gw17cp_post} shows the posterior distribution resulting from our first Bayesian updating step, the complete analysis of the GW170817 event. We showcase some relevant source parameters, both directly sampled (chirp mass, mass ratio and EoS index), as well as derived parameters (component masses, tidal deformabilities and radii). Our results are well within the expected and previously confirmed range for the event, see ref.~\cite{LIGOScientific:2018cki}.

\section{Full posterior of the agnostic prior} 
\label{App:Agn Posterior}

\begin{figure*}
    \centering
    \includegraphics[width=\textwidth]{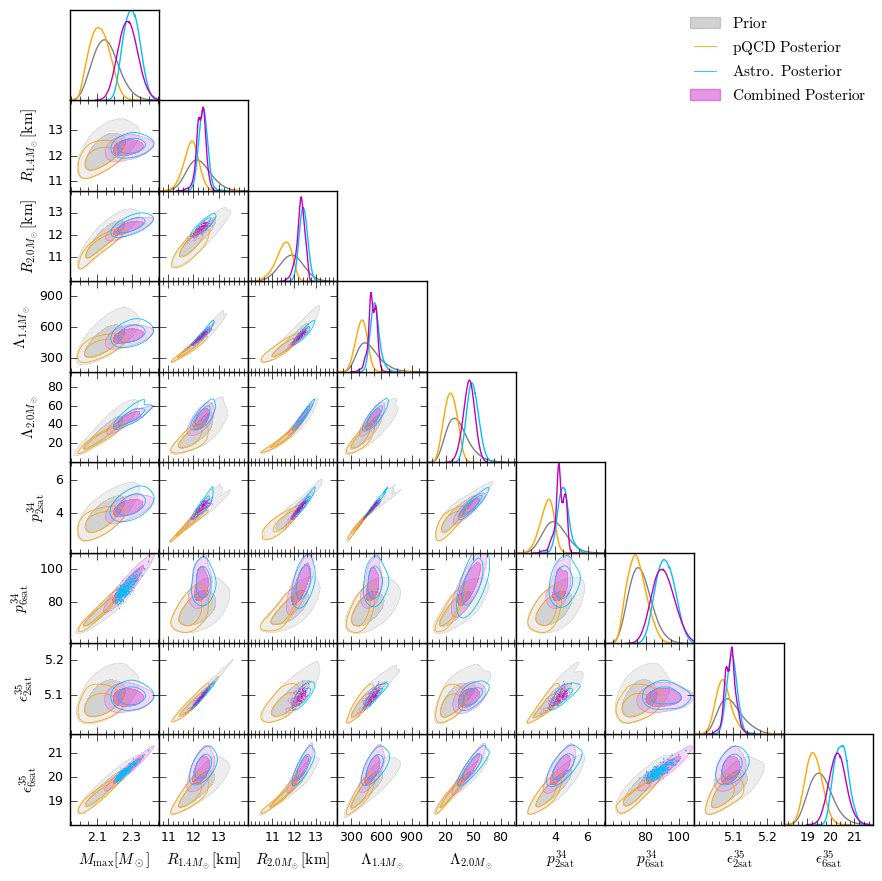}
    \caption{Corner plot of  macroscopic (maximum supported mass, predicted radius, and tidal deformability of a $1.4M_\odot$ and $2M_\odot$ NS) and microscopic (pressure and energy density at twice and six times nuclear saturation density) EoS properties. Prior is shown in gray, posterior after inclusion of astrophysical observations in light blue, pQCD calculations in orange, and the combined posterior (astrophysical and pQCD) in magenta. Pressure and energy densities at multiples of the saturation density are given in units of $10^{34}$ and $10^{35}{\rm dyne/cm^2}$, respectively.}
    \label{fig:cp_agn_post}
\end{figure*}

\begin{figure}
    \centering
    \includegraphics[width=\columnwidth]{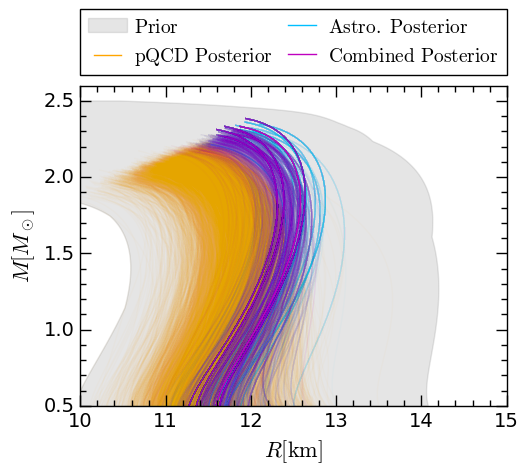}
    \caption{$95\%$ confidence interval of the posterior EoS distribution after including astrophysical data (light blue), pQCD constraints, (orange), and the combined result (magenta) in the mass-radius plane. The model agnostic prior is the gray background.}
    \label{fig:MR-post-agn}
\end{figure}

For completeness, we include a corner plot, Figure~\ref{fig:cp_agn_post}, of the full inferred parameters of a given EoS in the model agnostic prior, and the data posterior distributions after each Bayesian updating step, GW170817, X-ray and radio observations, as well as pQCD. As before, the two contours correspond to $68\%$ and $95\%$ confidence levels. The mass-radius posteriors and priors for the model agnostic GMM prior are shown in Figure~\ref{fig:MR-post-agn}.

\section{Effect of the pQCD  likelihood}
\label{App:pQCD}

\begin{figure}
    \centering
    \includegraphics[width=\columnwidth]{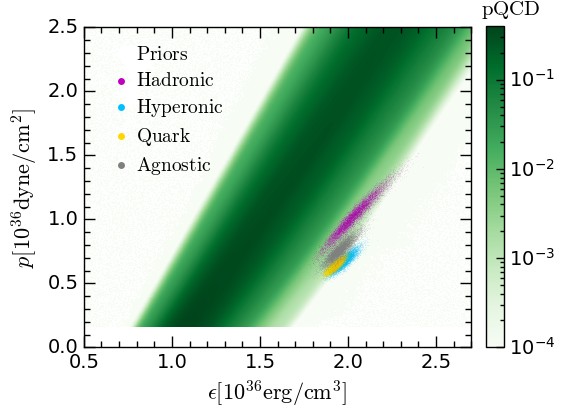}
    \caption{Prior marginalized pQCD likelihood (green colormap) introduced in ref.~\cite{Komoltsev:2023zor}, computed at $6\,\nsat$, for a range of pressures and energy densities. We overlay our model informed, hadronic (magenta), hyperonic (light blue) and quark (yellow), as well as the model agnostic priors (gray) predicted at $\epsilon(6\nsat),\,p(6\nsat)$, to showcase the domination of the differences between the priors over the likelihood.} 
    \label{fig:pQCDLike}
\end{figure}

The prior marginalized likelihood introduced in ref.~\cite{Komoltsev:2023zor}, unlike the original likelihood that joined the pQCD input to the lower NS density regime \cite{Gorda:2022jvk}, does not suffer from such a high dependency on the termination density, $n_\text{term}$. For a proof of how this likelihood constitutes a marginalization over the high-density prior, see  Appendix A of ref.~\cite{Komoltsev:2023zor}.

In line with the requirement that the pQCD likelihood should be computed at the highest number density studied and to prevent overly optimistic constraints, we have chosen to take the means of our prior distribution as $n_\text{term}$. For the model agnostic prior, $n_\text{term}=6.32\,\nsat$, while for the model informed prior, $n_\text{term}=6.25 \,\nsat$, $n_\text{term}=5.89\, \nsat$, and $n_\text{term}=6.09\,\nsat$ are the means of the number density  distributions of the hadronic, hyperonic and quark prior distribution, respectively. This leads to the minimally constrained/prior dominated posterior, when using the pQCD likelihood alone.

The effect of the pQCD input, in general, is to lower pressures at higher densities. 
The strength of this effect is heavily based on the relation between $n_\text{term}$ and any given EoS's $n_\text{TOV}$, the highest number density achievable at the end of the mass integration grid of the TOV equations.

We showcase the effect of the prior marginalized likelihood in Fig. \ref{fig:pQCDLike}. The green colormap represents the likelihood computed at $6\,\nsat$ for a range of pressure-energy density values. It is immediately obvious that across the main high-likelihood band, smaller pressures for a given energy density, and vice versa are preferred, as is expected to maintain causality beyond $n_\text{term}$, see e.g. ref.~\cite{Komoltsev:2021jzg}. Our model-informed priors are overlaid: hadronic (magenta), hyperonic (light blue) and quark (yellow), while the model agnostic prior is shown in grey. Although the aforementioned likelihood effect is observed on our priors as well, (lower left region of the pressure-energy density plane for a given number density is preferred), this effect pales in comparison to the difference between priors. Namely, the quark prior is highly localized and therefore can barely be constrained, and the hadronic prior as a whole predicts lower energy densities at a given pressure in the $\sim 4 - 6\,\nsat$ regime. We note this, in order to stress that, at least in this work, the observed pQCD prior marginalized likelihood effect, in this conservative scheme, is almost irrelevant when accounting for the prior differences.

\bibliographystyle{apsrev4-2}
\bibliography{references}

\end{document}